# Technical Report: **A Needs Assessment for Measuring Geographic - Legislative Associations in the U.S. House of Representatives**

Georgia Institute of Technology

Report Authors: Ashu Gupta, Cameron Owens, Benjamin Wilson-Langman, Max Hill, Jason Cox, Joshua Clinton, Clio Andris*

Research Team: Ashu Gupta, Cameron Owens, Benjamin Wilson-Langman, Max Hill, Jason Cox, Ian Arnold, Santiago Velez, Quinn Pearson, Yunhe Cui, Isaac Felix, Ryan Luong, Priya Parikh, Shubham Purish, Vivek Rao, Cat Tran, Mingkuan Yan, Christopher Yang, Clio Andris

Author Affiliations: Joshua Clinton is affiliated with the Department of Political Science at Vanderbilt University. All other authors and researchers are affiliated with Georgia Institute of Technology College of Computing.

*Corresponding author: Clio Andris, School of City & Regional Planning, School of Interactive Computing, Georgia Institute of Technology. Atlanta, Georgia. Email: clio@gatech.edu.

## Abstract

Political legislation affects the well-being and livelihoods of constituents. In the U.S. Congress a representative's voting record on bills and legislation is public. These bills have themes associated with them, such as veterans' affairs, coastal monitoring, agricultural appropriations, etc. A bill on veterans' affairs may affect a constituency differently if they have a high percentage of veterans. In this work, we demonstrate how congressional vote outcomes can be merged with typical geographic information systems (GIS) data to help compare a legislator's votes with the geographies of their constituencies to measure the association between district features and legislators' decisions.

We retrieved and tagged bills from the 118th U.S. House of Representatives (Jan. 2023 – Jan. 2025) by manually assigning each bill a set of themes. We then retrieved spatial data at the congressional district level related to each theme. We built a backend database to connect bill information and spatial data, and an interactive map prototype displayed the spatial data related to each theme associated with each bill. Our work helps identify challenges to creating a more accessible and complete system that would facilitate the visualization and analysis of the nature of contemporary Congressional representation. This technical report is primarily written for members of the data science community, and specifically those interested in legislative and geographic data.

A hyperlink to the map is available here: https://doi.org/10.6084/m9.figshare.32680776.

Keywords: Legislation, U.S. Congress, Congressional Districts, Roll Call Votes, Constituency, Geographic Information Science (GIS), Political Science, Geography, Information Visualization.

*AI Statement: No text in this document was generated using LLM technology.*

# Technical Report Summary

**Problem Statement**
The United States House of Representatives, a body of 435 voting members elected from geographic-based districts, introduces and votes on legislation encompassing a wide range of domestic and international issues. As an institution, the House of Representatives was intended to be the political institution that would reflect the diversity of interests, interests that were thought to vary based on geographic considerations (*Federalis*t 56). But how well does the voting behavior of contemporary representatives reflect the circumstances of the district they are elected to represent? Although we may expect differences in a representative republic with complicated and intertwined issues (e.g., *a representative from a district with a high percentage of workers in the oil and gas industry may vote either for or against legislation to strengthen worker safety measures in the oil and gas industry depending on the issue and interests)* we are primarily interested in systematically characterizing the extent to which the voting behavior of legislators can be associated with the characteristics of the district they represent.

**Work Performed**
This project explores how to link geographic information systems (GIS) data representing a Congressional district's characteristics with how that district's representative votes on legislation corresponding to those characteristics, and a method for visualizing this data for the public.

We retrieved and tagged hundreds of 118th Congress bills from the free, open U.S. Congress API. We used human-tagging, where we manually assigned each bill a set of 'keywords' (also called tags or themes) such as "potholes" or "veterans" using a data dictionary that we created. We then retrieved spatial data at the congressional district level that matched the theme. We built a back end to connect text files (.csv) (bill information) and GeoJSON files (spatial data).

Additionally, we built a front-end interactive map system that shows Congressional district polygons colored by a theme (e.g., percent of households on disability). The system also has an interactive box-and-whisker type chart and a T-test that distinguishes the districts whose legislators supported the bill from those that did not. It can tell us whether legislators who support a bill have different district characteristics than those who did not. It shows us who voted for or against a bill conditional on district characteristics.

**Theoretical Finding**
There are three primary methods for linking bills to GIS data: (1) site-specific, (2) thematic, and (3) keyword related. The site-specific method is useful for laws and projects with a clear geographically bounded area but requires more advanced GIS skills. The keyword method retrieves too many bills that are not relevant to a locale and omits relevant bills. In what follows, we use the thematic method.

**Results, Lessons Learned, and Suggestions**

- Omnibus bills, i.e., those with provisions across themes, are particularly challenging given the ambiguity of what it means to support or oppose them.

- Ceremonial bills (e.g., to name a post office) often have uniformly high agreement rates and arguably contribute little to our understanding of the nature of geographic representation.

- There is ample GIS data to match themes and it is often difficult to choose which of the many possible indicators (data variables) to use. A bill that supports *farm subsidies*, for example, could be described using measures based on *farm GDP, number of farms, percent of workers in agriculture, agricultural expenditures, change in agricultural land over time*, etc.

- Relevant GIS data that is aggregated to administrative boundaries is frequently found at the county level (such as infant mortality rates, HIV rates, or obesity rates). While county-level data is helpful and it can be projected into congressional districts with assumptions, it cannot be used in many places (e.g., downtown Los Angeles) because the districts are smaller than the counties or the boundaries do not align, despite prior research suggesting its use.

- Congressional district boundaries change frequently, as does the available spatial data, and elected representatives. Defining a narrow temporal scope, or a near-real time scope may be beneficial.

- Users may benefit from software that automatically suggests the GIS data themes that best match each bill but allows users to override suggestions.

- The methods described in this document can scale to local, state, and international government bodies.

# Technical Report: **A Needs Assessment for Measuring Geographic - Legislative Association in the U.S. House of Representatives**

## 1) Introduction

### 1.1 U.S. Political Representation and Geographic Space

The U.S. Congress is composed of two legislative chambers. The first chamber, the House of Representatives, elects new legislators every two years and is composed of 435 voting members each representing one of 435 contiguous geographic districts (plus 6 non-voting delegates). The number of districts in each state is based on the apportionment process following each decennial census and each state is constitutionally permitted to decide the procedures by which contiguous districts within each state are drawn. Contemporary districts must be similarly-sized within each state, and the average district now contains roughly 761,000 people (U.S. Census, 2020). The second chamber of Congress, the U.S. Senate, is also defined by geography, as each state elects two officials for staggered six-year terms (for a total of 100 members).

Elections for both the House and Senate are largely, but not exclusively, driven by partisan considerations. The role of political parties in organizing legislative behavior and shaping how voters interpret the political environment is well established. A two-party system simplifies an array of ideologies and can be used as a cue for voters to help them decide whom and what to support. Otherwise, it is difficult for constituents to monitor their legislator's voting record in Congress and assess whether their representative is helping or hurting district interests, as they understand these interests. To be clear, some interests are theoretical and/or normative (i.e., what district should prioritize is not always objective), but there are also pragmatic considerations associated with support or dissent for legislative activity. Even if constituents sought to track the actions of their elected representative or better understand the characteristics of the congressional district in which they reside, it is not easy for them to do either – to say about the difficulty of exploring the relationship between them.

*Legislative Activity* A roll call vote is a recorded legislative action (as opposed to voice votes, which are not recorded) that may be used by legislators to publicly note the positions of each member on legislation being advanced through the chamber (and on procedural issues related to the functioning of the chamber). Although roll call votes are not comprehensive measures of political activity, they are a key primary data source for legislative analysis (Vandoren, 1990). Roll call votes have been used to construct ideological models of Congress, including the landmark DW-NOMINATE metric, and to measure vote partisanship, historical trends in political activity, and ideological blocs within Congress (Jakulin et al., 2009; Poole and Rosenthal, 1985; 1991). Roll call votes are increasingly made based on political party lines. As journalist Caroline Paczkowski has written: "Over time, Congress has shifted from using roll call votes to document meaningful decisions — like passing bills or debating key amendments — to using them as messaging tools…These votes are not about governance; they're about optics" (Paczkowski, 2025).

***Constituency Needs & Assets*** The features and needs of congressional districts are broadly defined as the set of social, economic, and political issues that affect a congressional district, and for which we might expect district leaders or political parties to advocate for and legislate toward (Miller and Stokes, 1963). They include social, infrastructural, and physical assets such as national parks, bridges, festivals, and universities, and characteristics of the people and geography such as food insecurity levels, unemployment, education attainment, and extreme weather events, as well as the needs that follow these characteristics (e.g., more jobs or food assistance).

Unlike constituency opinions, which represent the political *views* of district constituents on topics such as allocations for university scholarships and gun ownership (Democracy Fund + UCLA, n.d.), our measure of *district need* is based on objective congressional district demographic and geographic characteristics, such as forest coverage, and agricultural output, among others. Although district preferences have been defined in terms of economic interests or demographic qualities, we lack a consistent method for mapping demographic or geographic features onto policy preferences (or legislation in general) (Adler, Cayton, and Griffin, 2018; Adler and Lapinski, 1997; Bogart and Vandoren, 1993). At the same time, a rich tapestry of public data on the assets and characteristics of geographic constituencies remains largely unconnected to questions of congressional representation. Surfacing these connections is important because although polls and surveys can be used to measure district opinion, these measures are often aggregated to provide an overall orientation rather than as a response to bill-specific opinions. Thus, it is unclear whether polling on specific issues can be treated as a policy demand. Furthermore, surveys and polls only represent a small sample of a constituency and can be expensive and cumbersome to administer.

### 1.2 Legislative Congruence

How a representative votes on bills relative to constituency characteristics (needs and assets) is a question of pervasive interest (Table 1). To be clear, representatives may seemingly vote contrary to constituents' needs for a variety of reasons (e.g., personal ideology or preference, or a lack of salience regarding needs) and it can be difficult to interpret how district features can reliably represent district preferences and need. In the first example, votes for H.R. 3738, the Veterans' Economic Opportunity and Transition Administration Act, differ without respect to percent veteran population (Table 1). In the second example, votes for H.R. 1, known colloquially as the Big Beautiful Bill, differ along party lines, despite both districts having similarly few households receiving benefits from the Supplemental Nutrition Assistance Program (SNAP).

Table 1: Examples of representatives, constituency characteristics, and votes for or against proposed legislation.

| District Rep. | District Characteristics | Voted For: |
|---|---|---|
| New York-07 (D) | 1.2% of constituents aged 18+ are veterans (lowest in the U.S., average: 6.04%) | **Increase in veterans' benefits *(H.R. 3738) Veterans' Economic Opportunity and Transition Administration Act (2024).*** |
| Texas-21 (R) | 10.7% of constituents aged 18+ are veterans (top 5th percentile) | **No increase in veterans' benefits.** |
| Kansas-03 (D) | 2.7% of constituents on SNAP (lowest in the U.S., average: 12.3%) | **Continued SNAP funding *(H.R. 1) The Big Beautiful Bill (2025).*** |
| Georgia-06 (R) | 3.1% of constituents on SNAP (second lowest in U.S.) | **Reduction in SNAP funding.** |
| Colorado-03 (R) (Lauren Boebert) | Fifteen (15) wildfire disaster declarations, as recorded by FEMA (2017-2022). | **No passage of *(H.R. 1066) Wildfire Recovery Act (2022).*** |
| Arizona-07 (D); California-18 (D); California-19 (D); Washington-02 (D) | Other districts with exactly 15 wildfire disaster declarations each, as recorded by FEMA (2017-2022). | **More wildfire recovery appropriations.** |

This discussion extends the long-established theory of *legislative congruence*, which can be defined as the alignment between roll call votes and the livelihoods and assets of society (Adams 1980, Hill and Hurley 1999, Miller and Stokes 1963). This term has been generally used to measure the (mis)alignment between legislators' voting behavior and the public opinion of their constituencies (e.g., constituents voice dissatisfaction with prohibition, but their representative votes in favor of it, indicating high legislative incongruence). Here, we used the term *legislative-geographic congruence* to instead capture the (mis)alignment between voting behavior and the documented characteristics of a constituency (e.g., a constituency is home to one of the U.S.'s 63 national parks, but its representative votes against a bill supporting national parks, indicating high legislative-geographic incongruence). Data detailing the characteristics of constituencies is widely available, although not presently consolidated, and in this work, we integrate that data to link a variety of topics to legislative outcomes. While prior work has variously connected legislative data (in the form of roll calls and amendments) to congressional district demographics (McDonagh, 1993), we aim to examine a more extensive set of district characteristics - involving infrastructure, employment, energy reliance, social services, and natural features, among others - with regard to congressional roll call votes.

**1.3 Research Objectives**

We aim to create a system that compares how legislators vote alongside a unique fingerprint of district characteristics that are presumably related to district preferences and needs. In so doing, we use trusted government data sources instead of polling and stated needs. Our goal is to build a data infrastructure that allows users to discover how legislative voting relates to district features in ways that may suggest a possible alignment score for rating votes and legislator-district pairings. If the system is used, the expected outcome is increased public knowledge about the needs of one's community and how these needs are being served. We have two research objectives:

1.3.1 Data Retrieval and Matching: We explain the types of data that are needed for this project and challenges to fusing and relating these datasets. These data structures could include spatial GIS data, node-edge network data, tabular data, temporal data and textual data, and thus present a data analysis and visualization challenge. The major data sources include bill text, legislator names and their votes on bills, and spatial data at the congressional district level. All data is from publicly available sources, and most data is hosted by government websites. We describe a model for a backend architecture and our experiences building the architecture.

1.3.2 Data Visualization: We outline a user-centered interactive map visualization that integrates legislator behavior and district data. The interactive map should allow users to see legislators' votes and district characteristics and should use common visual variables and information visualization techniques that are known to effectively help people engage with data considering best practices for interaction. We seek to learn about (a) what the tool enables (what can be done now but could not be done before) (b) common user tasks.

1.4 Anticipated Outcomes

Although this technical report does not include formal system deployment or user testing, it is designed to respond to the following problems.

***More Precise Measure of Congruence*** We currently lack methods that align legislator votes and constituency needs. By matching voting themes with district characteristics, we can better assess the levels of legislative congruence between constituents and representatives. Data science has not been leveraged to help find a solution to the theoretical loss of fidelity when representing districts. Doing so can combat misrepresentation and partisan influence and bring legislators closer to a more direct model of representation (Lascher, Kelman, and Kane, 1993).

***Constituent Disengagement and Dissatisfaction*** Constituents often do not know how their representative is voting on legislation, nor how their representatives votes are aligned or misaligned with their districts' needs and assets (also called "demand side measures" (Adler and Lapinski, 1997)). As such, it is unclear how a district will benefit from legislation, and whether legislative action is motivated by a district's specific needs as per the *instructed-delegate* model of governance or simply motivated by party allegiance as per the *responsible-party model* (Miller and Stokes, 1963). Some constituents who are aware of legislative actions have become increasingly dissatisfied with how their legislators are voting. This may be due to an "expectations gap" between constituents' needs and their legislators' voting behavior (Corbett, 2015). Increased political engagement, greater understanding of politicians' voting history among constituents, and a greater connection between politicians and constituents are proposed solutions to the expectations gap (Corbett, 2015). Providing constituents with more information about their district can help them assess their community's priorities.

***Combatting the Red/Blue Distinction*** Political polarization dominates the U.S. legislative landscape, and a legislator's political party is the clearest mechanism for predicting how a legislator will vote (Schoch and Brandes, 2020). Under this framework for political representation, termed the 'responsible-party model,' the unique, multifaceted sets of constituency needs are narrowed into largely binary sets of needs, as advocated for by legislators generally tied to their parties' platforms (Miller and Stokes, 1963). Despite the diversity inherent to 435 congressional districts — each with their own unique 'fingerprints' and with ample geographic data to numerically characterize these fingerprints — we lack reliable measurements to compare legislators' behavior with their districts' characteristics. Although representatives' partisan voting behaviors are generally aligned with their constituencies' partisan positions (Adams and Ferber, 1980), legislative constituencies are simply too varied and contain too many distinct factors to be adequately represented by the binary set of needs derived from party platforms. Providing data about districts from a wide variety of GIS sources can help differentiate districts beyond their political party representation.

This report proceeds as follows. Next, we describe the current landscape of elements that are needed for geographic-legislative congruence technology. Following, we explain proposed methods to reach our objectives. Finally, we share our proof of concept for a visual system and conclude with a discussion and suggestions for next steps.

# 2) State of the Art / State of Technology

In this section, we describe the state of the art / technology surrounding legislative data assimilation and visualization.

## 2.1 Legislative Data Access

The state of technology surrounding legislative data access and analysis is largely defined by two official channels for collecting legislative data. The first source of official legislative data is the U.S. House of Representatives' Office of the Clerk, which holds and displays records on roll call data, committee actions, and bills passing through the House of Representatives (Table 2). Congress.gov also contains roll call data, legislative actions, sponsorship records, and similar information online for both the House of Representatives and the Senate. The data stored in Congress.gov is easily accessible through the Congress.gov API, which allows for up to 5,000 requests per hour of bulk congressional data.

Most congressional data sources beyond Congress.gov and the Office of the Clerk are derived from those two and provide additional data packaging or analysis. Some of these third-party data sources include VoteView, GovTrack, Oklahoma University (OH)'s PIPC program, ProPublica's Congress API wrapper, and Sunlight Congress. The latter two are no longer fully supported but contain valuable data on past administrations and API maintenance. The former three - VoteView, GovTrack, and OH's PIPC program - contain roll call information in bulk, and both VoteView and GovTrack provide further analysis and visualization. Alternative sources for Congressional roll call data include the U.S. Office of the Clerk, Voteview, Govtrack, and C-SPAN. Historical votes from 1789-1998 can be found from the Inter-university Consortium for Political and Social Research (ICPSR), hosted by the University of Michigan.

## 2.2 Legislator Metrics in Data Analysis

Analysis of U.S. legislator behavior often includes legislators' relative positions in Congress, most commonly measuring their *ideological stance, partisanship score,* or *law-making influence* (Table 2). Measures of ideological stance are useful in models of congressional composition, with Poole and Rosenthal's NOMINATE and DW-NOMINATE scores driving their seminal work in modeling congressional ideology and partisanship over time (Poole and Rosenthal, 1991).

The Center for Effective Lawmaking (LawMakers, 2025), founded by the University of Virginia and Vanderbilt University, generates report cards for their representatives by calculating Legislative Effectiveness Scores, known as LES Scores. LES Scores constitute a measure of influence because they describe how often, how important, and to what extent a representative passes their legislation. Scores can be accessed using the requests library via a hyperlink to a spreadsheet (LawMakers, 2025). Other data sources such as Bridge Grades, which describes itself as "Rotten Tomatoes for politicians" (Bridge Grades, 2025), were excluded based on a lack of publicly available methodology descriptions or available methods but no description of data sources used.

Table 2: Online sources measure legislators' activity in a variety of metrics. Metric type is encoded by color: green for ideology metrics, blue for effectiveness metrics, orange for partisanship metrics, and white for alignment with the President. Citations to these sources are found in the references.

| Metric | Source | Description |
| --- | --- | --- |
| 1 DW-NOMINATE | Voteview | A dynamic, weighted three-step method to map legislators to two-dimensional **ideological** space by their roll call vote choices. |
| 2 Ideology Score | GovTrack | Positions legislators on a one-dimensional **ideological** scale by their bill cosponsorships. |
| 3 Cook Partisan Voting Index (PVI) | Cook Political Report | **Ideological** stance of congressional districts on a one-dimensional scale linked to political party alignment. Each district's metric is relative to the national mean. |
| 4 Leadership Score | GovTrack | Positions legislators to a leadership scale by the frequency with which they **co-sponsor** bills compared to other legislators. |
| 5 Legislative Effectiveness Score (LES) | Center for Effective Lawmaking | Measures the frequency with which legislators are able to accomplish **tasks**, including introducing bills, sponsoring bills, pushing them through Congress, and making them into law. |
| 6 Congressional Rhetoric Metrics | Polarization Research Lab | Measures the frequency and content of legislators' **communication** across floor speeches, newsletters, press releases, and official social media accounts. |
| 7 Party Loyalty | Voteview | Percentage of the time each legislator votes with a **plurality** of their party, versus evenly split and against. |
| 8 Bipartisan Index | The Lugar Center | Measures the frequency with which legislators **co-sponsor** bills with legislators of other political parties. |
| 9 Presidential Support | Voteview | Fraction of legislators' votes in which they vote for the **President's** stance. |

**2.3 Data Visualization for Legislative Voting and Bills**

In everyday life, the prevailing method of visualizing the outcome of legislative votes is to simply list the legislators and their vote (Figure 1). Many examples of these lists are accessible online, often through individuals' social media accounts. In some cases, the output will be helpfully divided into yay and nay votes. Another common method of summarizing vote outcomes is to divide responses by party affiliation.

Key alternative resources include Voteview and GovTrack, which help users visualize roll call vote outcomes using dashboard-type visualizations at the bill level. Voteview (Lewis et al., 2021) represents legislators as points along two axes, each representing a dimension defined using PCA and defining predicted voting spaces using a high-dimensional vector (Figure 2). Voteview also uses a choropleth map of districts with bivariate colors to highlight party agreements and deviations in legislator votes. GovTrack provides *bar graphs and a unit-based visualization* of a congressional floor layout to show yay and nay votes.

In visual systems research, *bill topics* have been used to identify the ideological leanings of legislators and we use the theme approach in this research. Academic researchers have also examined methods of visualizing legislative data Many Bills (Aktolga et al., 2011; Assogba et al., 2011) which ran publicly from 2010-2012, allowed users to search through bills, as represented graphically with squares upon a timeline and colored by topic (also described in DocBlocks (Assogba et al., 2010)). The SensePlace data visualization suite (MacEachren et al., 2011, Pezanowski et al., 2018) integrated geopolitical text documents with a map by searching for place names in the text in real time.

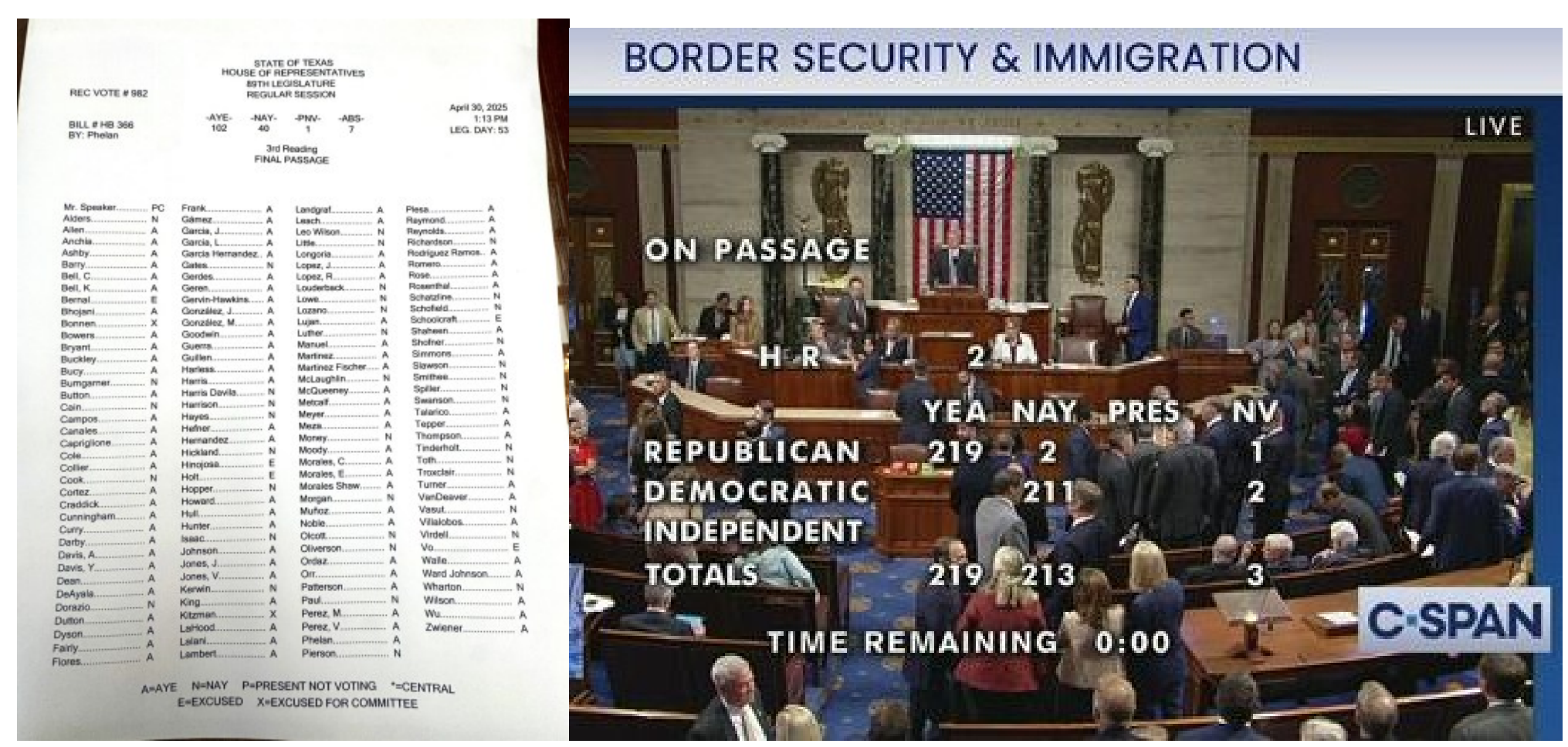

Figure 1: Legislative vote outcomes are often presented in reductive ways. **Left**: A sheet of papers lists legislators and their votes for a 2025 Texas State House of Representatives bill, uploaded by Rep. Brian Harrison (R-TX-HD 10) and shared on X (Harrison, 2025). **Right**: A TV capture of a U.S. House of Representatives vote on a 2023 border security and immigration bill, shared by Rep. Chellie Pingree (D-ME-1) (Pingree, 2023) shows the number of representatives in each party who voted for, against, a bill or did not vote.

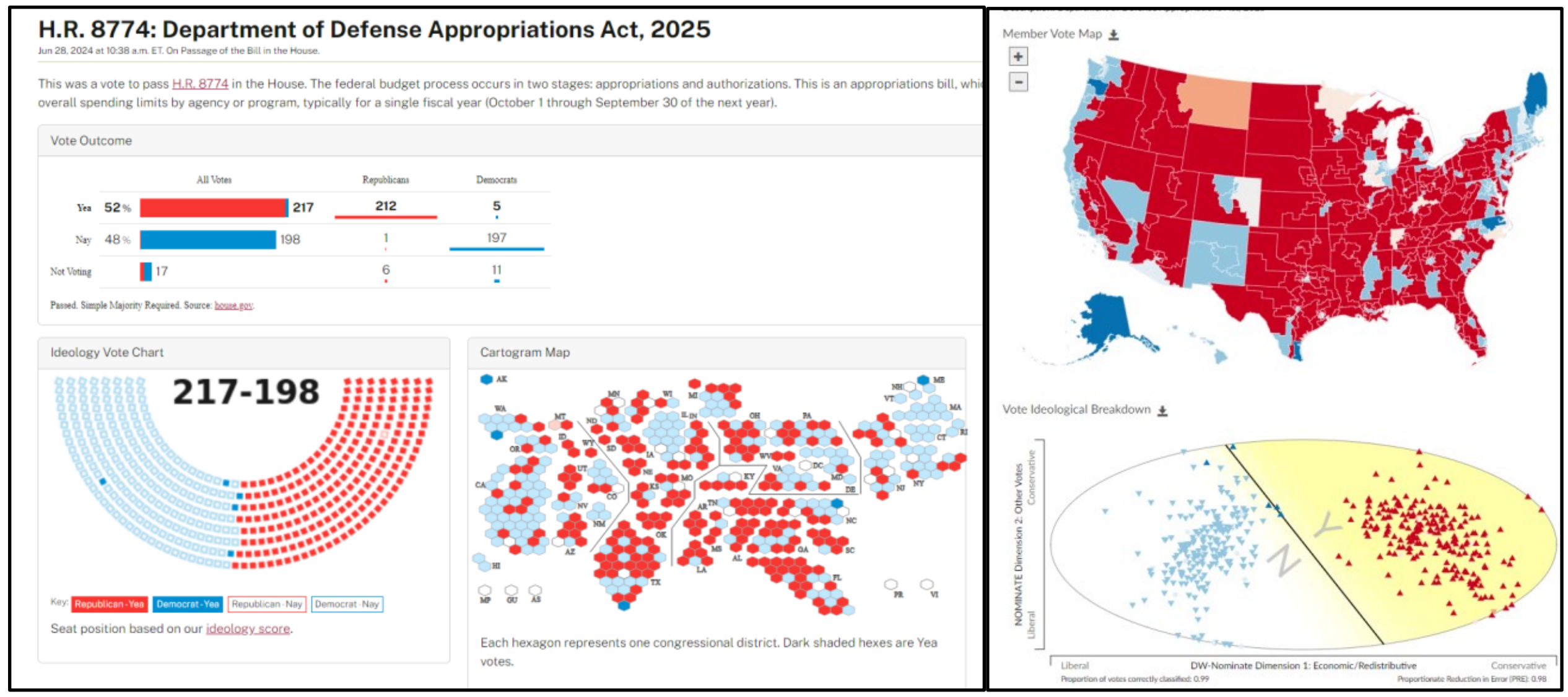

Figure 2: Visualizations from Voteview (L) and Govtrack (R), show the outcomes of US H.R.8774: On Passage of the Department of Defense Appropriations Act, 2025. The bar chart, glyph, maps and feature reduction techniques are initial examples that help individuals interact with legislative data.

# 3) Data Processing Approaches

**3.1 Approaches for Linking Geographic Data to Bills**

There is a lack of technical guidance on how to link legislative documents to geographic space. In this work we choose to use the thematic data approach to link bills to geographic space, as discussed in the previous section. However, we initially considered three different approaches, as outlined below (Table 3).

***Thematic Approach*** To infer locational data values from the topic of the bill (e.g., a bill about veterans is matched with data on the number of veterans in a district), we **match the topic to thematic data.** This method allows us to consider the widest variety of bills, as we do not need a bill to have a specific site or place name. It also provides variation of continuous values across districts–for instance, the percent of forest coverage per district, for which the other methods are not well-situated. However, common issues of temporality, fidelity, sources, and geospatial entity aggregation may still arise.

***Keyword Approach*** We can conduct a simple **keyword** search, where we retrieve place names that are listed within a bill. The congress attempts to do so by listing regions, countries, and states that are found in bill text. Users can search for bills by location on the congressional website (Congress.gov, 2026). While these methods are easy to computationally perform, the presence of a keyword does not always focus on the geographic unit it references. Additionally, district level locations are not well represented within keywords provided in most bills.

***Site Specific Approach*** In some cases, a bill is passed that **defines a specific locale** (e.g., legislation arguing for Chesapeake Bay Clean Up) and thus it is relatively clear what geographic area will be impacted by this bill (Table 5). However, the areas surrounding a focal entity may also be affected and this may go unacknowledged, or the entity may be represented as a single point or series of points (e.g., two coal mines), and it is not clear which geographic polygonal area to designate as 'affected'. This is especially difficult when the bill does not map to a particular administrative unit but instead may refer to a watershed, a path, or a region (such as the Mississippi Delta).

Table 3: Three theoretical approaches for linking a legislative document to a geographic area.

| Approach for linking a bill to a district's geography | Example | Benefits | Challenges |
|---|---|---|---|
| (1) Thematic | A bill about veterans' education funding and geographic data on the number of veterans in a district. | -There is ample data on topics (e.g., forests, demographics, bridges). The data also are often continuous and can be used for statistical analysis (e.g., correlation).<br>-Data is often in easy-to-use GIS format.<br>-Data is relatively easy to understand using choropleth maps. | -Temporal mismatch between data and bill votes.<br><br>-Subjective, often fuzzy, matching of a bill's theme to a GIS dataset's theme. |
| (2) Keyword | The term Massachusetts is referenced 27 times in 118th Congress' legislation. | -Prevailing method used in Congress-dot-gov.<br>-Easy to compute and retrieve.<br>-Produces scalar integer values.<br>-Administrative boundaries are easy to attach to keywords. | -Appearance of a place name does not equate to place relevance (e.g., the "Norway Pact" as applied to a US policy).<br><br>-Many areas are not easily named, and -district-level foci are not well represented. |
| (3) Site Specific | Allocation of funding for: Superior National Forest, Hudson River Valley, C&O Canal, Alabama Underwater Forest, Lahaina Heritage Area. | -A specific site location is described in detail. Bounded geographies are relatively easy to represent.<br><br>-Often the co-sponsors are from the affected districts. | -Imprecise areal delineation, variation on buffer sizes (i.e., the areas nearby or connected that would be affected).<br><br>-Misinterpretation is possible, the official geographic region may be the Mississippi Delta when the bill only affects part of the Delta. |

**3.2 Measuring Legislator Support for a Topic**

There are multiple ways to measure a legislator's support for a topic. In many cases, it is difficult to synthesize the key issues at stake in a bill, and the priorities of legislators when they eventually cast their vote. A classic example is that support for the extraction of raw materials benefits a community because it provides jobs and economic benefits. However, this can also be harmful because of destruction to the environment and increase of pollutants (among other reasons). Thus, measuring support can be subjective. We draw on precedents to evaluate legislative support, such as report cards and media, and explain the costs and benefits of aggregating yay and nay votes on the same topic.

***Report Cards*** A method of combining data points to produce synthesized records. Below are two examples where associations provide A-F grades based on a legislator's voting record, co-sponsorship and other factors. The *National Education Association* provides legislator report cards. They use 14 different roll votes over the 118th Congress, on 11 bills (one bill had 4 roll call votes related to it and its amendments) as input for the scorecard, and these are chosen "based on their relevance to advancing NEA's identified legislative priorities" (NEA, 2025). The *NAACP* also puts out report card statistics (NAACP, 2025) with a percentage grade as well an A-F and used 32 roll call votes in the 115th Congress.

Report cards are beneficial because they communicate a grade from A-F, and this is a familiar standard for the average user. However, it is not always clear how the legislator received the grade, despite efforts to describe the process, and at some points, the grade can be subjective or reductive.

***Media and Press Releases*** Releases can be used to help explain a legislator's choice of vote and help researchers measure their support for or against an issue. For example, four Utah representatives explained why they voted for the H.R.-1 (The Big Beautiful Bill) in a 2025 press release (Kennedy et al., 2025). Legislators also post information explaining their (lack of) support for a bill on social media accounts and are quoted in news media stories. Researchers can extract themes (e.g., it provides more housing) from these types of texts and then compare the reasoning to actual data in these districts.

***Aggregating Yays and Nays by Thematic Bill Topic*** We use this method to summarize a legislator's support for a position. With any data on legislator behavior, it is not clear whether we should provide statistics that amass the outcomes of bills on a set of topics (e.g., Oil & Gas). For instance, to say that a representative voted YES on 4 out of 12 bills related to Oil & Gas is very helpful because it can allow us to see the representative's support for passing legislation on that issue. However, if some of the bills were beneficial for the industry, and some were harmful to the industry, the value of 4 out of 12 would need more refining to extract the representative's support or lack of support for Oil & Gas. This method allows the user to see how often a legislator voted for or against a topic, avoiding overemphasizing a single bill. However, it can be an oversimplification to combine many bills into one category because they may have considerations that need to be individually examined.

### 3.3 Approaches for Bill and Theme Retrieval

To compare a legislator's voting record to geographic characteristics and needs of their constituencies, we identified geographical and spatial themes linked to these characteristics and assigned each bill to associated themes. We describe the major challenges associated with this task.

***Codebook*** To identify geographical and spatial themes relevant to legislative constituencies, members from the research team developed a numeric codebook/data dictionary to organize themes and record GIS data sources (see Appendix B). This document's themes are listed roughly from physical data to behavioral data, which generally mirrors the temporal development geospatial data from the 1960s to the present.

***Tagging Process*** Bills were then 'tagged' with appropriate theme keywords. This was done manually using the theme keyword list and the Congress.gov website to view bills. The team had a shared web-based spreadsheet with each bill's information for the 118th Congress. Each row had the bill title, roll call vote tally, a tag, and a short description. They populated a column called "Tag" with their impression of each bill's main topic (Table 4), and these tags were cross-referenced by other group members.

Table 4: A data snippet with applied theme tags. Colors are provided as a general, subjective way of categorizing tags into higher level themes.

| **Bill** | **Year** | **Name** | **Subject (Tag)** |
|---|---|---|---|
| H.R.2225 | 2021 | National Science Foundation for the Future Act | Technology |
| H.R.8958 | 2024 | NASA Reauthorization Act of 2024 | NASA |
| H.R.2811 | 2023 | Limit, Save, Grow Act of 2023 | Energy |
| H.R.3684 | 2021 | Infrastructure Investment and Jobs Act | Rural Development |
| H.R.1319 | 2021 | American Rescue Plan Act of 2021 | Rural Development |
| H.R.1339 | 2023 | Precision Agriculture Satellite Connectivity Act | Agriculture |
| H.R.615 | 2024 | Protecting Access for Hunters and Anglers Act of 2023 | Hunting |
| H.R.8790 | 2024 | Fix Our Forests Act | Forestry |
| H.R.1147 | 2023 | Whole Milk for Healthy Kids Act of 2023 | Nutrition |
| H.R.3617 | 2022 | MORE Act | Cannabis |

The bill type, bill topic, and special notes were manually added to each entry. Bill codes were available for some bills via Voteview (Lewis et al., 2021), with a tradeoff: many bills were tagged with a high-level code (specifically, Clausen and Peltzman codes) but these codes did not provide detail about the bill. Conversely, the Congressional Research Service (CRS) Policy Area and Issue Codes (Poole and Rosenthal, 1991) provided more detail, but few bills had these tags (see Appendix A).

***Retrieving Bill Data*** We chose to include the following subjects: NASA, Domestic Oil and Gas, U.S. Veterans, Wildfire Prevention and Recovery, Flooding and Tropical Storm Prevention and Recovery, and Domestic Agriculture in our search. Once a filtered list of bills was returned, we manually searched the returned bills to determine whether each bill is both relevant and useful. This process follows some guiding questions:

- Is the bill related to this topic, and this topic alone? In other words, could a legislator's behavior regarding this bill be attributed to some other subject?
- Is the bill substantive enough to reflect the legislator's position or behavior regarding the topic? In other words, would the legislators' positions be communicated through this bill?
- Could it be argued that both the YEAs and the NAYs support the topic, just from different ideological perspectives – or is one group of voters in support and one against? If the latter, the bill is more useful in determining stance.

# 4) Proof of Concept Static Maps: Wildfires and Agriculture

We developed two proof-of-concept static maps and performed basic analysis on topics related to physical geography (wildfires) and economic geography (agriculture). We chose these topics because they are highly linked to land (physical space) and livelihoods, and they are not often described as politically charged issues (compared with immigration, reproductive rights, etc.). The goal of these studies is to demonstrate the linkage between votes and district characteristics and to point out flaws that arise in the bill-geography matching process.

## 4.1 Wildfire Frequency

*Do legislators with many wildfires in their districts vote for wildfire protections?* Wildfires are increasingly frequent within the United States, especially in California, Oregon, and Washington. Legislators have attempted to address both potential causes and negative effects of the wildfires through legislation. The 118th Congress voted on three bills supporting protections from wildfires (Table 5, Figure 3).

There is little evidence connecting fire prevalence within a congressional district to support wildfire bills (Pearson Correlation Coefficient of 0.05). Most representatives supported the bills, with each representative voting "Yea" for an average of 2.33 out of 3 bills (Figure 3). Support for wildfire bills is mostly non-partisan: on average, Republican representatives voted for 0.44 more bills than Democratic representatives (2.55 to 2.1). The only partisan bill was the *Fix Our Forest Ac*t, with only 25.9% (55 out of 212) of Democratic representatives voting "Yea" and 96.8% of Republicans voting "Yea". Doug LaMalfa (CA-1 R), Jared Huffman's (CA-2 D), and Vince Fong (CA-20 R) did not support every wildfire bill but had 105, 56, and 55 fires in their districts, respectively. (Note that LaMalfa and Fong did not vote on all three bills). Wildfire data is sourced from FEMA (2024).

***Takeaway*** The wildfire proof of concept case study focuses on legislators who did not vote for bills, although two such legislators were not present for the votes. This case raises discussion about whether support should be measured as a yay vote, or a lack of nay votes (e.g., not voting).

### 4.2 Agriculture

*Do legislators with high agricultural output vote for bills supporting agriculture?* Agriculture is a large, multi-billion-dollar industry, with 174.9 billion dollars of agricultural products exported according to The American Farm Bureau Federation. The most valuable farm markets within the United States are concentrated in the Great Plains and Midwest region, with many congressional districts in states such as Nebraska, Iowa, Kansas, North Dakota, South Dakota, and Minnesota having sold over 10 billion dollars of product on the market in 2022 (USDA Agricultural Census, 2022). We analyzed roll call votes of three bills related to agriculture in the 118th Congress (Table 6, Figure 4).

The market value of the products sold in a congressional district loosely correlates with "Yea" votes for agriculture bills (0.28 Pearson correlation coefficient). On average, republican representatives voted for one more bill than democrat representatives (2.87 to 1.76). Republican Kelly Armstrong (ND-0) did not vote for one bill even though their district's market value exceeds $10 Billion. Additionally, twelve representatives actively voted against at least one of the bills. Republican Matt Gaetz (FL-1) voted against two bills, while most other legislators who did not support bills were absent for the votes.

***Takeaway*** The agriculture proof of concept case study illustrates that bills we retrieved with agricultural keywords were perhaps only loosely related to agriculture. Even if the bills were more closely related to agriculture, it is unclear whether supporting these bills equated with supporting U.S. agriculture. Thus, the relationship between support and large agricultural output is subjective.

Table 5: Three bills voted on in the 118th Congress related to wildfire protections.

| **Title** | **Overall vote; Yays \| Nays \| Not Voting (NV)** | **Description** |
|---|---|---|
| *H.R. 4866* (2024) Fire Weather Development Act of 2024 | **341-51**<br>D: Yea: 191 Nay: 0 NV: 21<br>R: Yea: 150 Nay: 48 NV: 18 | Establishes multiple programs and requirements aimed at addressing wildfire forecasting, detection, and management, as well as Interagency Coordinating Committee on Wildfires and the National Advisory Committee on Wildfires. |
| *H.R. 8790* (2024) Fix Our Forests Act | **266-148**<br>D: Yea: 55 Nay: 151 NV: 6<br>R: Yea: 213 Nay: 0 NV: 7 | Establishes new requirements for reducing wildfire projects while expediting and exempting certain activities from NEPA review. It limits consultation requirements related to threatened species and legal litigation in relation to fireshed management. It implements projects and technologies to reduce wildfire risks while promoting forest restoration and stewardship. |
| *H.R. 1567* (2023) ACRES Act | **151-258**<br>D: Yea: 200 Nay: 0 NV: 12<br>R: Yea: 206 Nay: 4 NV: 11 | Requires the Dept. of Agriculture and Dept. of the Interior to submit reports on the reduction of hazardous fuels. This relates specifically to vegetation management activity to limit wildfire risks. |

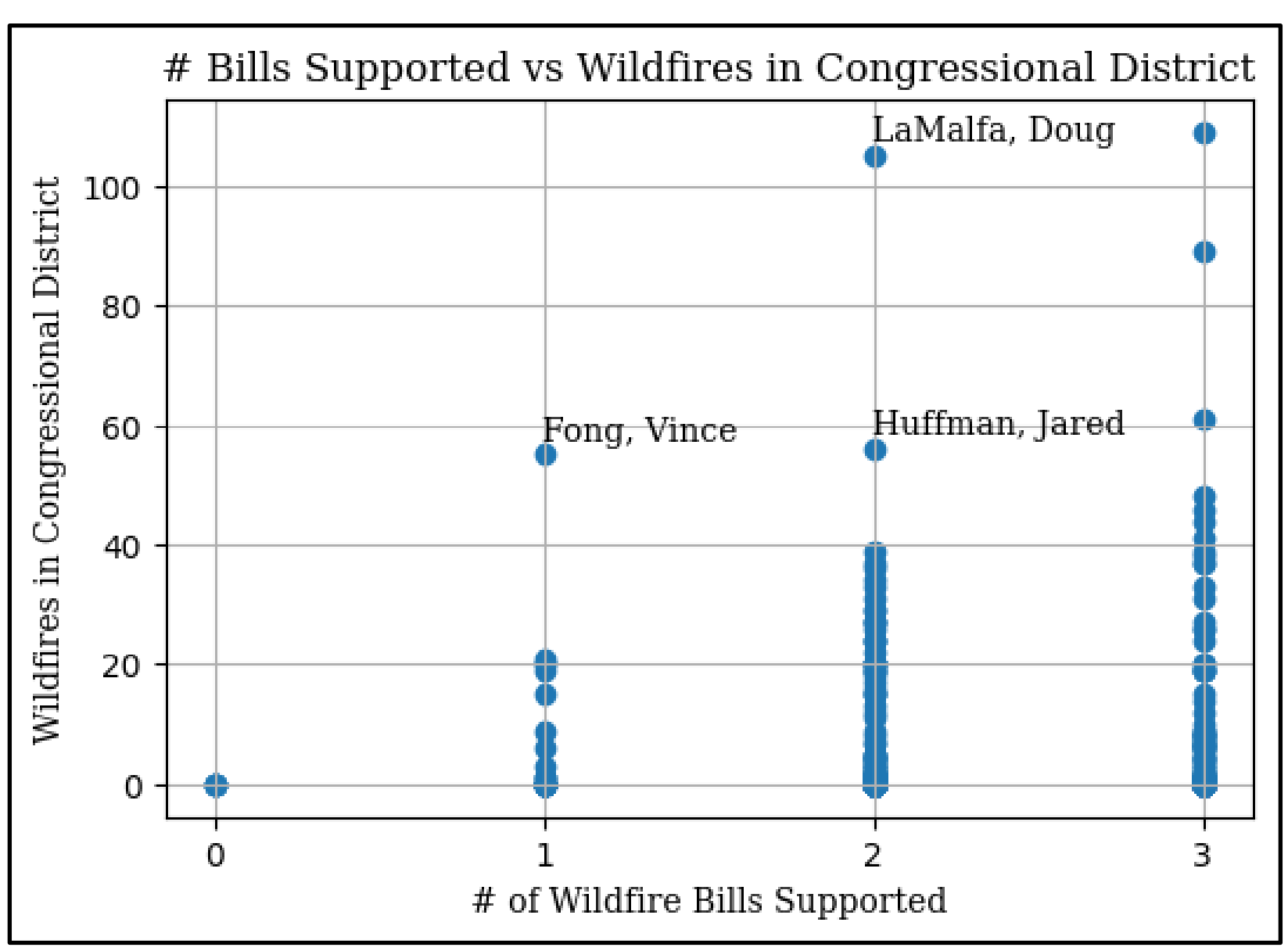

Figure 3a: A scatterplot shows the number of wildfires within each congressional district and the number of wildfire bills their legislator supported. Each dot is a representative / district. Representatives Fong, LaMalfa, and Huffman are distinguished with labels.

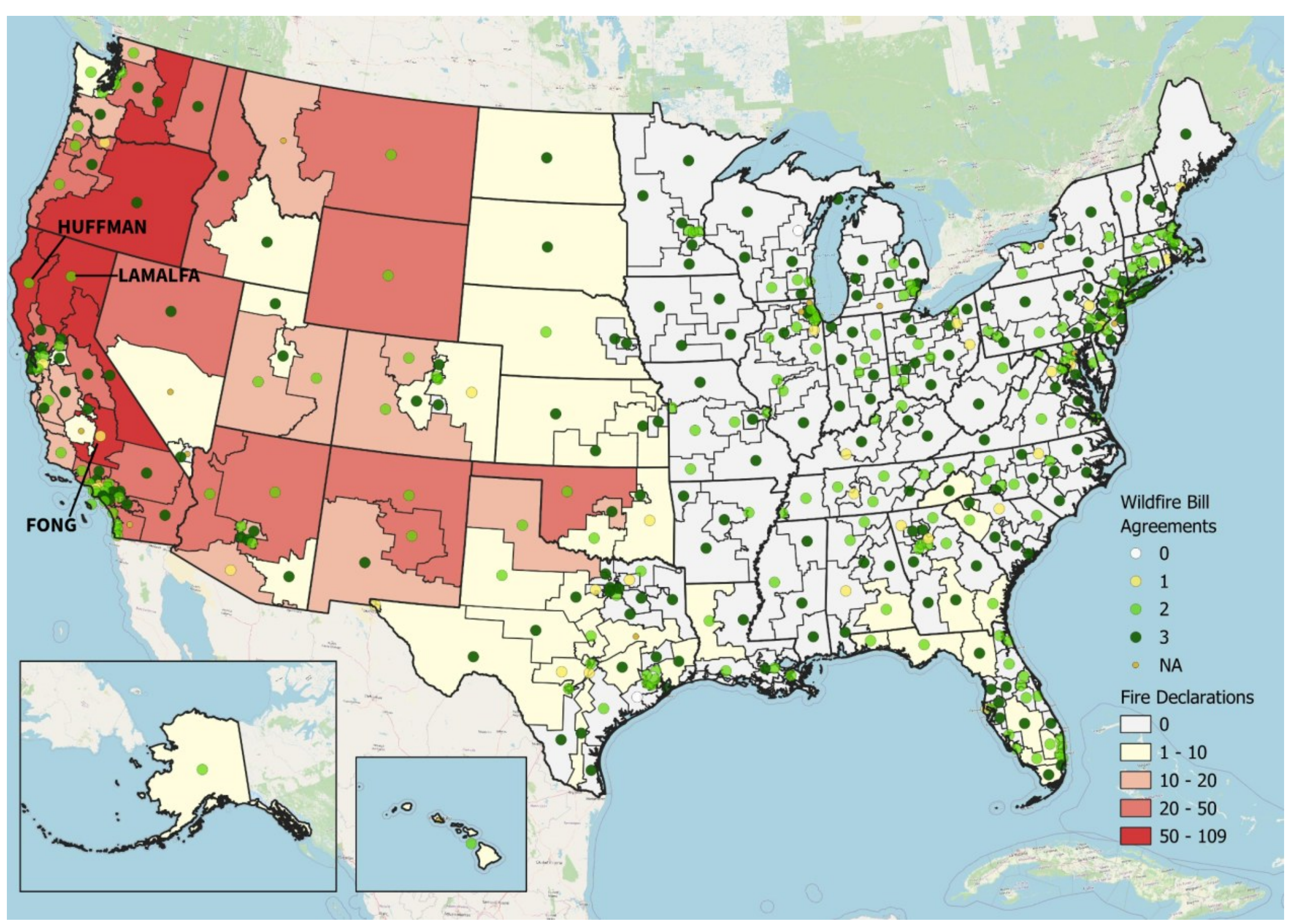

Figure 3b: A map shows fire declarations vs. support for three fire protection bills. Representatives Fong, LaMalfa, and Huffman are distinguished with labels.

Table 6. Three bills voted on in the 118th Congress related to agriculture.

| Title | Yays \| Nays \| Not Voting (NV) | Description |
|---|---|---|
| H.R. 9456 (2024) *Protecting American Agriculture from Foreign Adversaries Act of 2024* | D: Yea: 55 Nay: 148 NV: 8<br>R: Yea: 214 Nay: 1 NV: 4 | Expands the powers of the Committee on Foreign Investment to determine whether, for a land transaction, a national security review or another action needs to be taken because of foreign investment risks. |
| H.R. 1339 (2023) *Precision Agriculture Satellite Connectivity Act* | D: Yea: 207 Nay: 0 NV: 6<br>R: Yea: 202 Nay: 11 NV: 9 | Orders FCC to review and recommend changes to its regulations for the use of satellites for precision agriculture. |
| H.R. 1147 (2023) *Whole Milk for Healthy Kids Act of 2023* | D: Yea: 112 Nay: 98 NV: 3<br>R: Yea: 218 Nay: 1 NV: 1 | Modifies the national school lunch program such that whole and reduced-fat milk options are offered. Milkfat would be excluded from restrictions on the saturated fat in meals. |

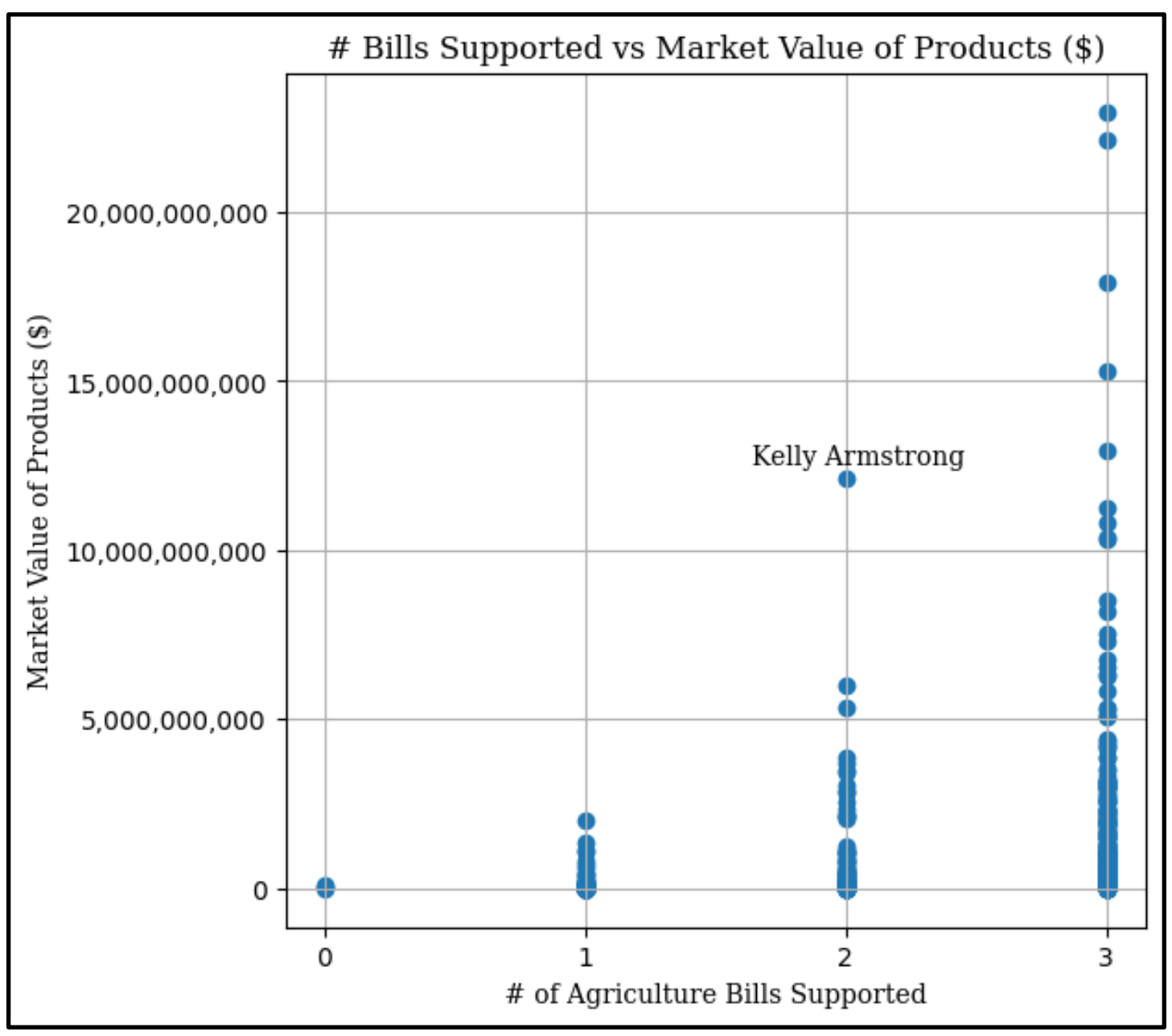

Figure 4a: The scatterplot shows the market value of products within each congressional district and the number of agriculture bills the corresponding legislator supported. Each dot is a representative / district.

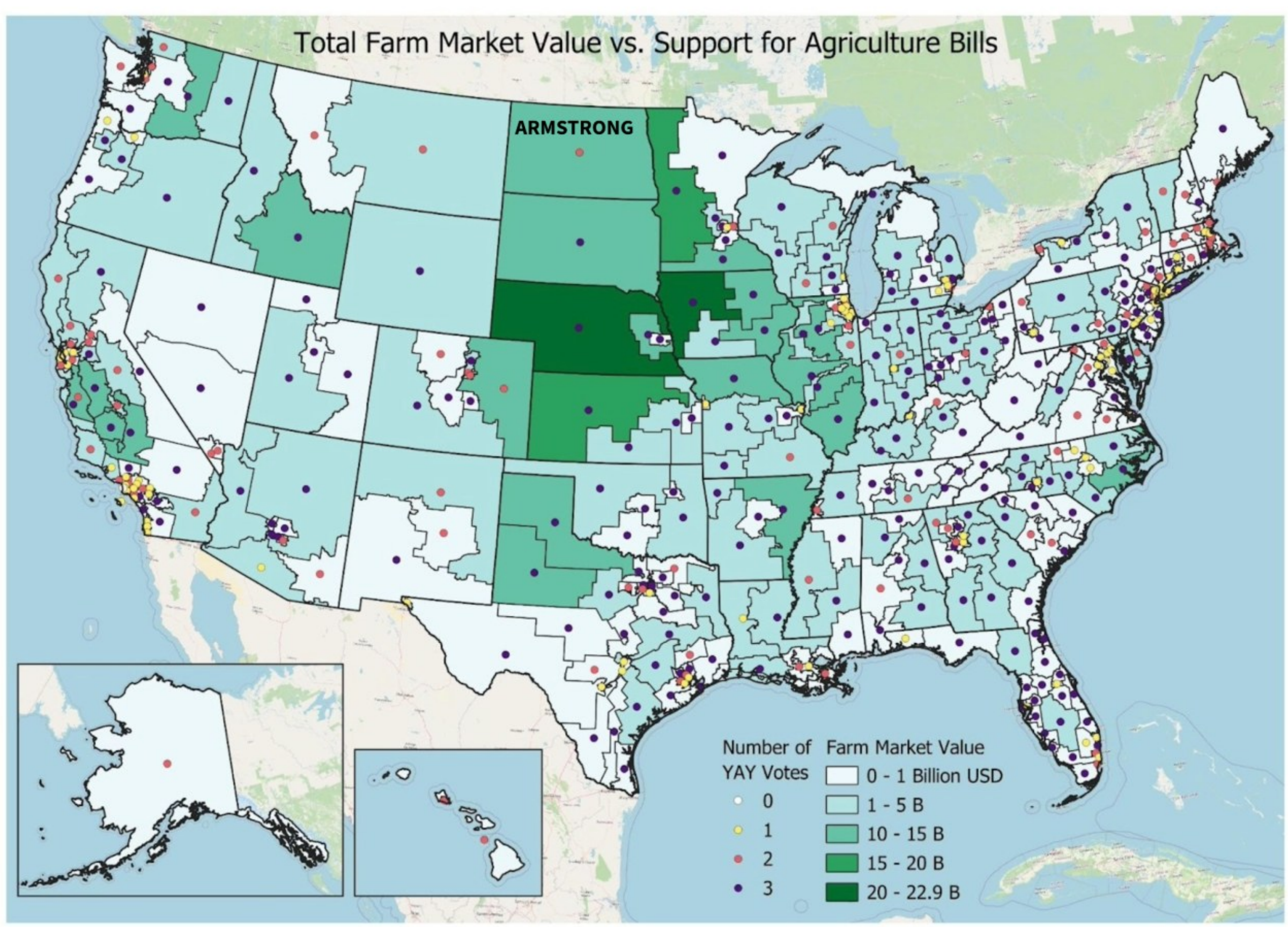

Figure 4b: A map shows farm market value vs. support for three agricultural bills.

# 5) Technology Stack, Data Architecture, and Data Pipeline

This section describes the methods used to implement a scalable, interactive legislative visualization. It documents the technology stack, data architecture, and data pipeline used to ingest, integrate, and display district geometries, vote outcomes, and congressional district metrics.

**5.1 Technology Stack**
The interactive visualization system is implemented as a web application to reduce access barriers (no account nor installation needed). The website is built using a modern React and TypeScript technology stack, selected to balance performance, maintainability, and reproducibility. The stack can broadly be divided into two groups: core technologies and development tools. In combination, these groups support the project's key technical requirements: fast interactivity, linked visualization across multiple coordinated views, consistent handling of heterogeneous datasets, and a maintainable codebase that can be extended as additional data and visualizations modes are added. The rationale behind each component selection is described in further detail below.

**5.1.1 Core Technologies** The core technologies define the applications interaction model, data flow, and rendering behavior. These determine the function and feel that the user experiences.

***React UI Library*** React provides a component-based user interface (UI) model well-suited to composing the application's distinct interactive elements (map, plot, selectors, controls, tooltips and statistical views). React Router enables client-side navigation between views without full page reloads. React hooks serve as the foundation for triggering data and component updates following user interaction.

***TypeScript Programming Language*** TypeScript's enforcement of explicit data shapes helps reduce error when integrating heterogeneous datasets and improving refactor safety as the application evolves.

***D3 Visualization Library*** Data Driven Documents (D3) is used as a visualization utility layer that provides scales, axes, and brush selection for the plot component, and cartographic utilities for the map component (Albers USA projection, path generation, and choropleth color scales). D3's zoom behavior enables pan and zoom navigation of district geometries.

***Zustand State Management*** Zustand enables lightweight client-side data stores with predictable state updates and minimal boilerplate. This facilitates cross-feature interaction, such as the linked views between the map and plot, by allowing user focus events (hover, click, brush selections) to be written once to a shared state and consumed by multiple components.

***Bootstrap CSS Framework*** Bootstrap provides consistent UI styling and accessibility-friendly controls across screen sizes. Bootstrap's standardized components reduce custom CSS overhead, expediting control surface development (dropdowns, drawers, navigation) while maintaining a coherent visual style.

**5.1.2 Development Tools** The development tools do not directly impact the user experience. Instead, they benefit iterative development workflow through enhancing code reliability and maintainability.

***GitHub Repository, Version Control, and CI/CD*** GitHub is used for source control, collaborative development, automated validation, and automated deployment. A branch-based workflow supports reviewable changes. Continuous integration (CI) and continuous deployment (CD) pipelines are used to execute verification of code quality then deploy code to its hosting location.

***Vite Build System*** Vite is a modern frontend tool that provides speedy hot module replacement (HMR) during development and optimized production bundling during build. These features were advantageous to the visualization-heavy development cycle.

***Node Package Manager (NPM) & Runtime*** Node provides the JavaScript runtime used for development and build tooling. It enables the Vite developer server and production build pipeline, supports dependency management via NPM, and runs automation scripts (linting, formatting, testing).

***Quality Tooling: ESLint, Prettier, and Vitest*** Code quality is enforced using ESLint and Prettier to keep the codebase consistent and readable across contributors. Automated tests are implemented with Vitest to validate core data parsing and cross-domain identifiers. This is particularly important in the data pipeline as silent mismatches when integrating multiple data sources can cause misleading visual outputs.

### 5.2 Data Architecture

The interactive system uses a static file-based data layer. Pre-processed JSON files are committed to the repository and fetched individually by the client on demand. Datasets are organized into three complementary domains: (1) Legislative, (2) Metric, and (3) Geographic (Figure 5). These are packaged with the site build to avoid the need for a database or API layer in the prototype, while keeping data organized, inspectable, and reproducible.

***Legislative Data Domain*** Legislative data is stored as one JSON object per bill, using a two-part structure: a metadata block and a data block. The metadata block includes bill information and readable descriptors needed for the UI. The data block is a dictionary keyed by district identifier. Each value records a district's vote outcome and associated legislator metrics. Legislator metrics are stored per-bill, instead of a congress-wide lookup. This accounts for legislator turnover, party alignment changes, and other edge cases within a congressional term. Vote outcomes are kept as raw strings in the data files and normalized at runtime to a small, enumerated set (yea, nay, no vote) so that visual components can group and aggregate data reliably.

***Metric Data Domain*** Metric data are stored as one JSON object per metric and follow the same overall pattern as bills: a metadata block describes the metric, and a data block contains per district values. The data block is a single-layered dictionary keyed by district identifier with numeric values (e.g., counts, percentages). This structure supports fast lookup during rendering and makes metrics interchangeable in the UI; values can be swapped while leaving the join logic unchanged.

***Geographic Data Domain*** District boundaries are represented as TopoJSON, which encode polygons as shared arcs so adjacent districts reuse boundary segments; this improves load time and reduces file size compared to raw polygon GeoJSON. At runtime, the client browser converts the TopoJSON topology into a GeoJSON Feature Collection for compatibility with standard mapping and D3 path/projection tooling. A companion geometry layer of district point features is stored in GeoJSON format and is used for centroid-based interactions such as hover targets, labels, or point overlays.

***District Entity Integration Layer*** All three data domains (legislative, metric, and geographic) share a canonical join key that also serves as a district identifier. This architecture is implemented as a static, file-based storage instead of a normalized relational database. We describe the system's interpreted structure as an entity relationship (ER) diagram. The ER diagram shows how the three data domains relate to one another through the unifying congressional district entity (Figure 5).

Each district is uniquely identified by a canonical district identifier derived from the congress number, state abbreviation, and district number. The legislative domain is joined with districts through congressional district vote records. The metric domain is joined with districts through per-district metric value records. The geographic domain provides district spatial representation. In implementation, the data is structured as domain specific nested JSON objects optimized for static hosting and on-demand loading.

### 5.3 Data Pipeline

The interactive system data pipeline features manually collecting upstream datasets, transforming data offline, then packaging information into the repository directory (Figure 6). During build, Vite copies these static assets into the production bundle, and the client fetches only the files needed for the user's current views. This design circumvents the need for a backend API for the interactive system prototype while still supporting responsive interaction.

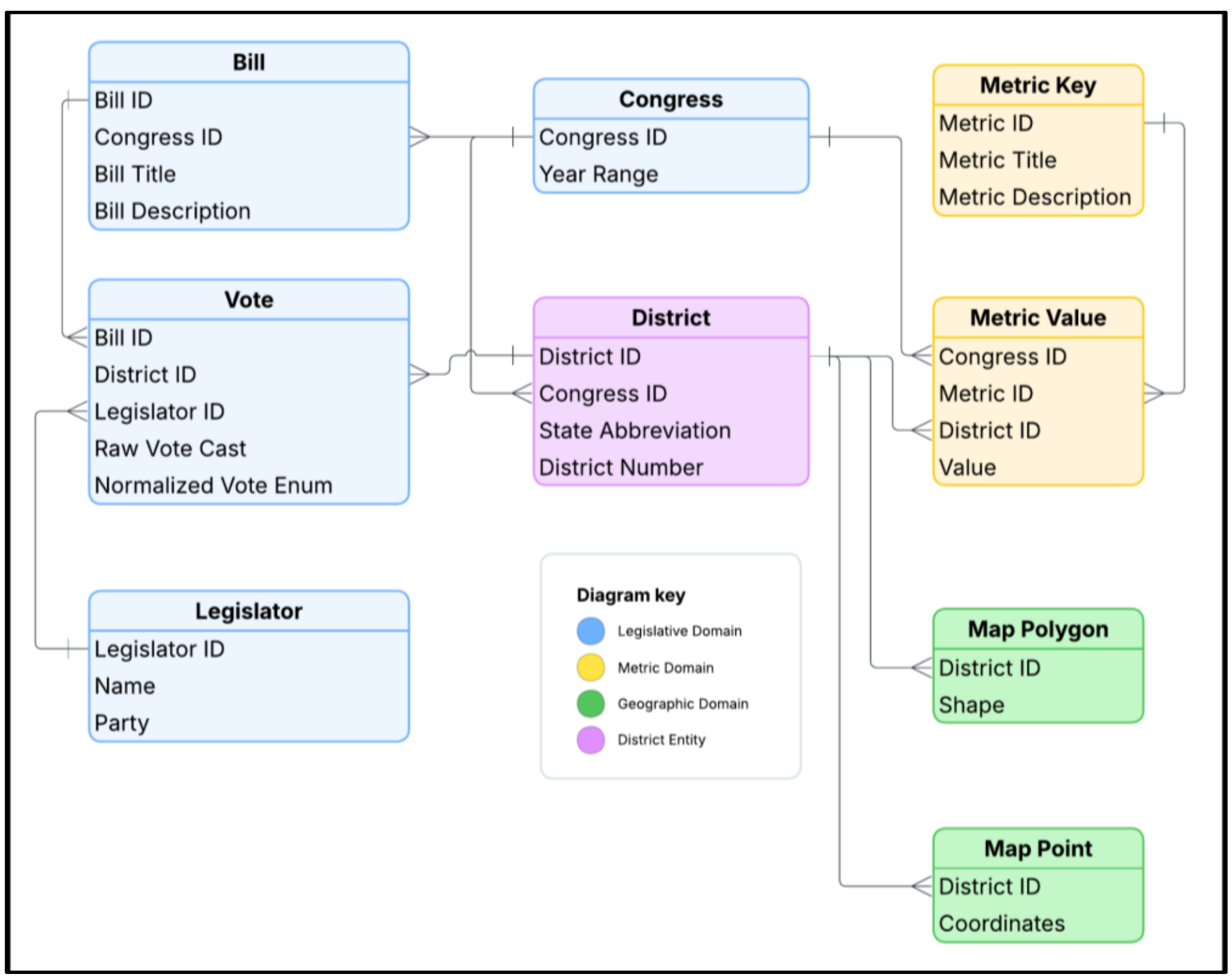

Figure 5: Legislative, metric, and geographic data are joined together via district identifier in an entity relationship diagram.

***Offline ETL Processing*** Each data domain follows the following extract, transform, and load (ETL) process. District polygons are downloaded, locally converted from shapefiles to TopoJSON format, then keyed by district identifiers. Polygon centroid points are generated from district polygon properties. Legislative data is pulled from disparate API sources, aggregated, and reshaped into the described one bill per file format. Metric data is similarly pulled from API sources, filtered, and reshaped into the one metric per file format. All processed data is then loaded into the app repository (Figure 6).

***Build-time Packaging and Indexing*** To keep the initial page load fast, the client does not download all bill and metric files at startup. Instead, the build process generates lightweight index files used to populate data selectors and discover available data.

***Browser Runtime*** At runtime, the app loads only what is necessary for the current view. On start up, index files are fetched, and default data selections are established. When the user changes a bill or metric selection, the corresponding JSON file is fetched, and its contents are written into a Zustand data store. The data store serves as the single source of truth for loaded districts, vote records, and metric values. Aggregate statistics (e.g., vote counts, metric averages) are recomputed within the store whenever the underlying data selection changes. A separate visualization Zustand store holds transient UI states (e.g., hovered districts, active selections, color settings), enabling linked interactions between the map and plot. App components subscribe to slices of these stores via selectors, so only the affected views are re-rendered when state values change.

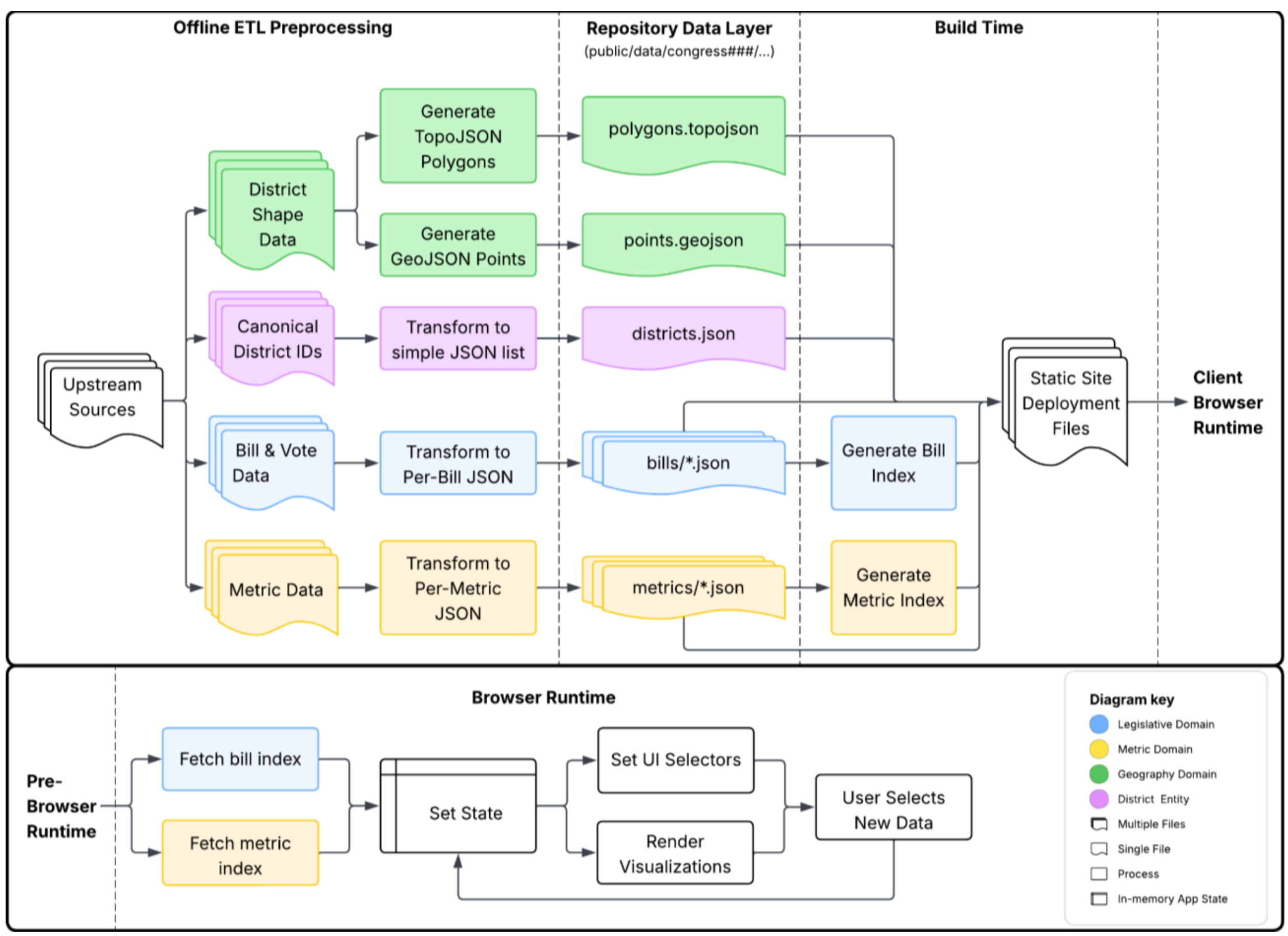

Figure 6: The interactive system data pipeline includes preprocessing for data extraction and transformation. Processed data is then saved to the system repository and hosted with the application, where it is made available to the client through state data stores.

# 6) Interactive Map System

### 6.1 Interactive System Needs

The needs for an interactive map are derived from HCI and GIS literature as well as other bespoke visualization systems that have helped users navigate congressional data (Draper and Riesenfeld, 2008; Grosvenor et al., 2013; Kalouli et al., 2018; Kinnaird et al., 2010; Silva et al., 2018). The system should be designed to be as easy to use as a common weather map. A system for comparing the alignment of bill vote outcomes with constituency characteristics must (1) link legislative behavior to relevant geographic characteristics of constituencies, (2) be scalable so that it can be replicated to local government (e.g., city council districts) and international levels in order to benefit a variety of stakeholders, (3) be transparent so that individuals can see how measures are formed, (4) allow users to read and select individual districts and legislator names, (5) provide numeric facts that help users assess alignment, (6) be accessible to a wide swath of the public with as few barriers as possible, (7) be sharable and `ownable' so that the public can create visualizations, and look up information to share with others virtually and in person, and use to learn about legislative decisions in their district and other districts. The system should use classic types of interactions: linking and brushing, interactive selection of map and plot(s), hovering, tooltips, dropdowns, and filters.

### 6.2 Interactive Map Description

Our prototype is available here: https://doi.org/10.6084/m9.figshare.32680776. It is a public, query-based tool that helps users access, explore, and visualize congruency between vote outcomes and physical, infrastructural, health, demographic, and energy-related GIS data metrics for the 118th U.S. House of Representatives.

***Bill and Metric Selection*** In the **Data Selection** module, users can select a Congress and then choose a bill and companion metric to visualize on the map (Figure 8). Users can expand a section entitled **Bill Description** to show a brief description of the bill's text. Bill descriptions vary in length and this bar is collapsed by default to minimize clutter. The user can modify the colors of the plots (both yea/nay votes and district fill color) in the options tab.

***Statistics*** Users can retrieve answers to simple questions such as: "Are legislators who voted against this bill representing districts with smaller veteran populations than expected?" We first show the reader the number of votes and average map feature data value (e.g., percent veterans) within a voting category (yea/nay/no vote). We also compare the average map data value per voting category to the entire distribution of map data values using a Welch's t-test for each bill-metric-vote pairing (via Python's SciPy library (SciPy, 2008)). If the test's alpha value is greater than 0.05, we display a "Yes" on the statistics panel to denote the significance of each metric-vote pairing in the summary component (Figure 7).

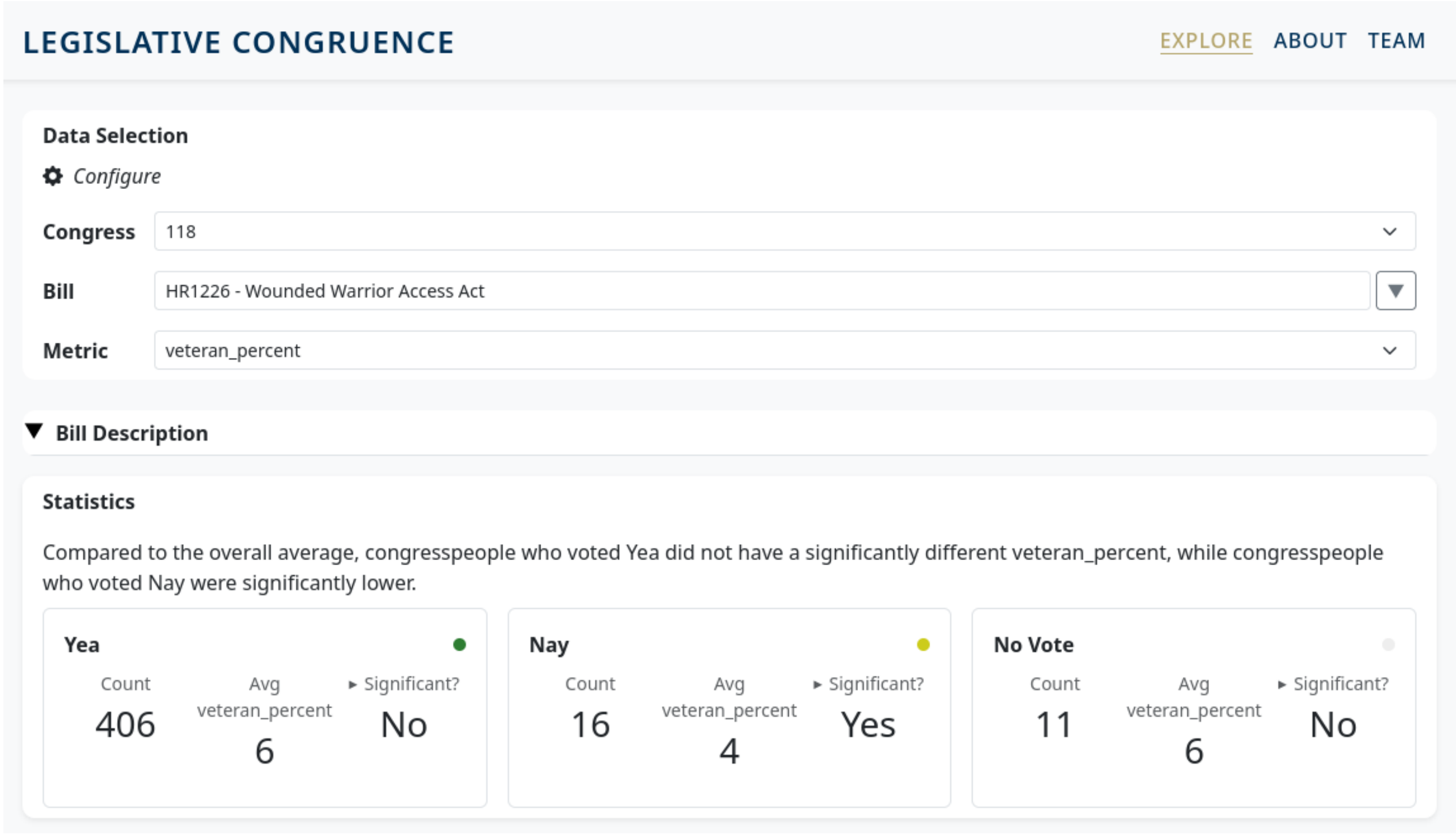

Figure 7: The selection components include **Data Selection**, **Bill Description** and **Statistics**. This example shows the average percentage of veterans in each district vs. how their legislator voted for the H.R. 1226 (Wounded Warrior Access Act).

***Map*** Once the selections are made, the resulting map shows congressional districts that are colored based on the metric, and points at the center of each district, colored by how the member voted. Users can hover over or click points to show a summary tooltip, including the legislator's vote, their district's value for the selected metric, and legislator information (Figures 8, 9). Points are colored to indicate legislators' votes (dark green for aye/yes/yea, yellow for nay/no, grey for no vote), and districts are colored in greyscale, with darker greys corresponding to higher values in the selected metric.

***Plot*** Next to the map is a plot of legislator votes (binary yea/nay) against the selected metric. A linked plot divides votes into yays and nays by color, shape, and position (dark green diamond for yea and yellow circle for nay). The plot shows **three dashed lines** at the mean of metric values under three voting category subsections (yea, nay, or no vote/absentee) (Figure 8) to help the user see which districts are above or below average. These lines also help the user see whether the average yay voter is from districts with much different characteristics than the average nay voter, or conversely, if yay and nay votes come from places with similar characteristics.

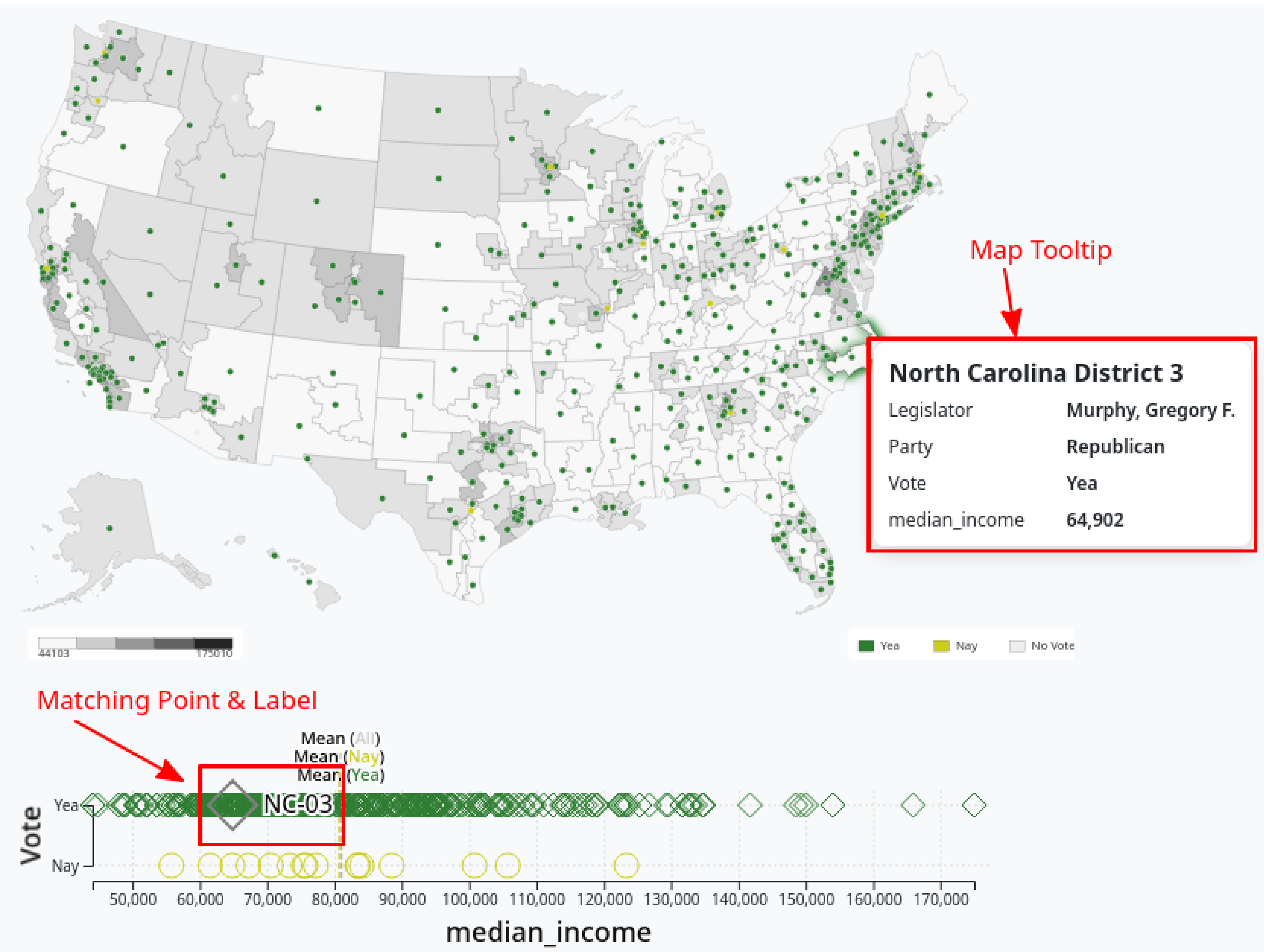

Figure 8: The interactive map shows districts colored by median income (greyscale) where darker colors indicate higher income. Dots represent legislators (and are placed at the center of each district) and colored by whether the legislators voted for (green) or against (yellow) H.R. 1226 (Wounded Warrior Access Act). North Carolina District 3 is highlighted on the map and on the plot.

***Interaction and Syncing*** Users can brush to select a point on the plot, and then the corresponding district(s) will be highlighted on the map. For example, a user can select all points whose districts have greater than 315,000 household vehicles, and the map will automatically highlight those districts (Figure 9).

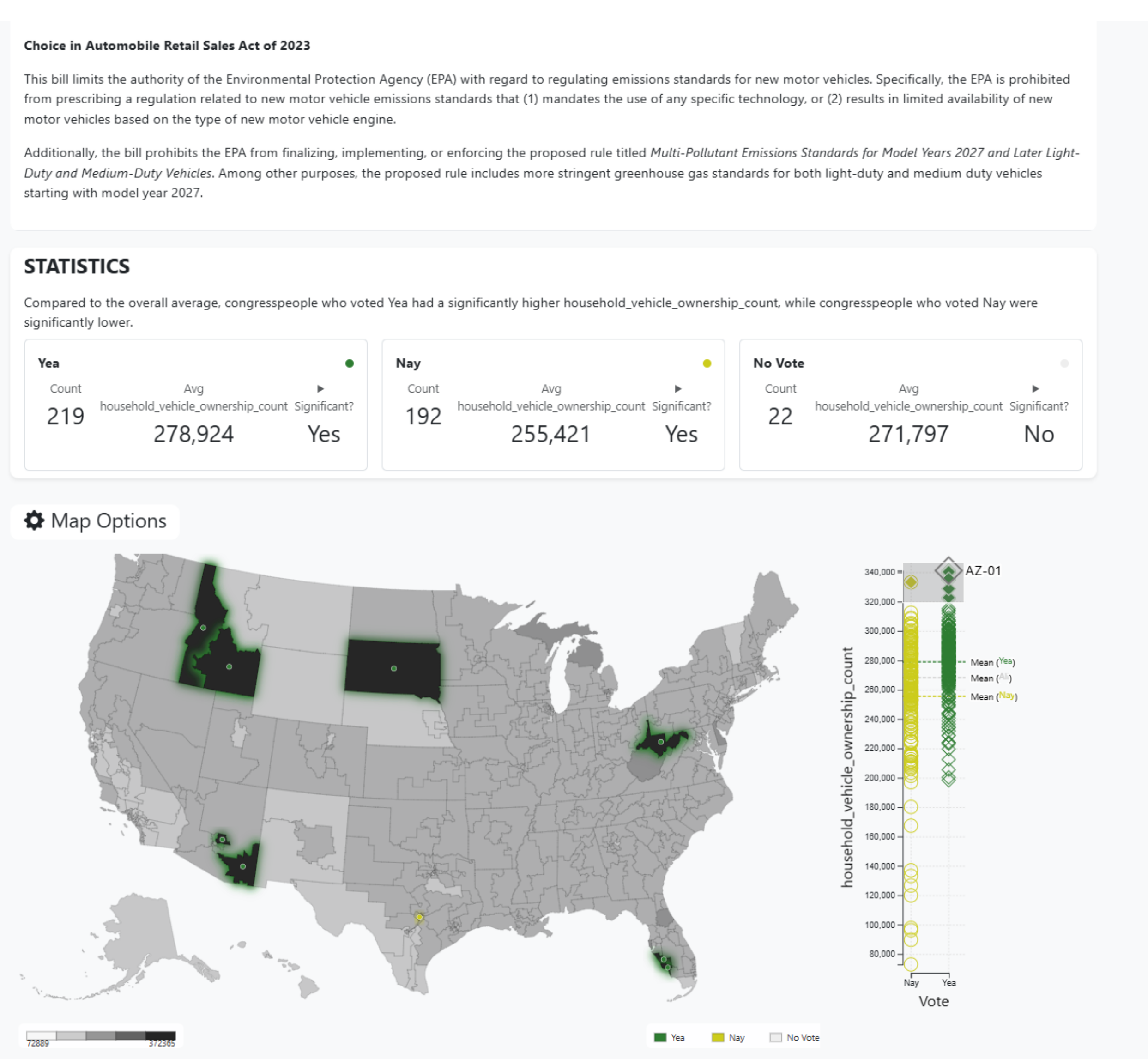

Figure 9: An example of a group of brushed (highlighted) points in the plot with their corresponding districts highlighted on the map. Each point represents one legislator (which also equates to one district). The selected districts have the most households with automobiles, and all but one legislator (in the Dallas, Texas area) voted for H.R. 4468 (Choice in Automobile Retail Sales Act of 2023).

The tool enables:
-Comparison of vote outcomes (yay/nay) with district characteristics
-Overall view of district characteristics
-Overall view of bill popularity
-View of a specific representative's behavior
-View of a potential correlation of the district characteristic and the vote outcome
-Variation of outcomes and characteristics with a state or region
-Visualization of similar behavior among neighboring locations, urban/rural divide, and coastal/inland divide.

# 7) Discussion and Conclusion

### 7.1 Lessons Learned

Regarding wider limitations, the implementation of measuring legislative congruence using large datasets has *many* limitations, caveats, and pitfalls. Foremostly, there can be incorrect linkages between GIS data theme and bill theme (e.g., matching household ownership rate with a bill on healthcare), which can result in spurious correlations or off topic analyses. Next, extracting themes from bills was difficult, as bills can have many clauses and provisions and voting one way or another on a bill may not reflect one's support for certain aspects of the bill. The data changes over time, not only with changes in legislators, but district boundaries, and versions of GIS data (Census data year to year, climate data, etc.). Temporal matching may require substantial effort and a large data team. Keeping the map up to date to show near real-time results would be helpful for users, and older votes may not be as valuable in a visualization system as votes that are emerging and relevant in the news cycle.

Furthermore, linking a bill to geographic space that implies causation is difficult. Therefore, it is challenging to know how constituents or places are helped or hindered by a piece of legislation. For instance: (a) a bill could affect one district at the time of its inception, but another in the future, (b) what appeared to be detrimental to an area may show benefits over time, (c) a bill can ignite future legislation that affects other areas, and (d) relationships can be made by sponsors and voters that may be visible only later on. Examining how past legislation may have affected an area may serve as a predictive model that could be applied to future scenarios.

### 7.2 Future Work

Future work may include hypothesis testing related to information gain, public opinion, and rates of sharing information. The technical report lacks voices from practitioners and researchers in the fields of political science and policy. Instead, it provided a geography / GIS and data science perspective on merging datasets and mapping.

***LLMs*** Large language models (LLMs) have been used in other projects to extract key information from state level bills and to help assess legislators' stance on topics (Davoodi and Goldwasser, 2024). LLMs can provide a fast alternative to hand-tagging bills, and in the future, automating the process of bill tagging using LLMs may be helpful. Future models should be trained and checked using a human-in-the-loop method where the human should examine the model outputs and assess whether its tags make sense, given the composition of the bill text.

***Backend Services*** The current interactive system relies on a client-side web application architecture that prioritizes rapid prototyping at the cost of long-term scalability that a dedicated database with API layer could provide. Developing backend infrastructure would allow the system to support larger and more diverse datasets while improving flexibility in how information is stored, queried, and served to the interactive system. A relational database would also strengthen the data pipeline by enforcing schema consistency, preserving relationships across entities and supporting more reliable data ingestion as new sources are added. For analysis workflows, this would make it easier to join datasets across domains, enabling faster exploration analysis and more efficient generation of derived statistics.

***Alternative Map and Data Representations*** Alternative representations to a traditional choropleth map include using an equal area hexagonal bin **cartogram** where each district's area is the same and correlate with their number of representatives. In addition, a tree map representation may effectively show yay/nay votes (encoded as color and position) alongside the districts' characteristics (encoded as area). This representation lacks location information, but allows the district's theme (e.g., GDP from tourism) to be prominently encoded (as area). The treemap method would also allow users to get quick, immediate information about the popularity of a vote. In addition, a traditional bivariate map would probably help reduce cognitive burden on the reader, as only a single polygon layer would be used, instead of dots atop polygons.

## 8) Conclusions

Legislation affects livelihoods and the environment, and currently, outcomes are presented to the public as lists of legislator names along with their party and state abbreviation, and their vote. Current methods of displaying the data require users to scroll through lists of hundreds of names and makes it difficult for readers to remember and engage with the outcomes. There are rarely any visual variables or encodings to help the reader. Next, it can be difficult for users to find their own representative or a specific individual from the list, without using a search function. Additionally, users can only see one vote outcome at a time, where they may want to look at small multiples or select themes on a common topic.

In response, we illustrated a method for aligning legislator votes with constituency / district characteristics by creating a data architecture that includes both roll call votes and GIS datasets, and an interactive map responding to key research questions about legislative congruence. We took a visual and GIS-approach to examine these issues by situating the legislator in the context of their geographic district.

The result is a way of looking at legislation to see how it may affect districts differently. It also draws the reader closer to constituency characteristics by displaying district features alongside the vote of the elected representative. In so doing, the system helps users better understand the nature of legislative activity and its' connection to geographically defined interests and communities. This can help enrich conversations about the nature of politics by providing another lens besides the typical partisan frame to help identify sources of commonality and similar circumstances that may recenter conversations about the people being represented rather than the politicians who are supposedly doing the representing.

To that end, we hope this research helps others better connect legislative voting by elected officials and constituent and district circumstances. By presenting an approach that can elevate information about member behavior, district characteristics, and the associations between them, we hope our efforts help citizens better understand the nature of their government and the actions of their elected officials.

# Appendix A

Table A: An overview of the types of codes, the amount of bills each covered by each code type, and notes about code types gathered during the bill tagging process.

| **Code Type:** | Coverage Percentage: | Notes: |
|---|---|---|
| **Clausen Codes** | 91% Coverage | 6 unique codes used to categorize bill type (not bill theme) via Clausen (1973). |
| **Peltzman Codes** | 91% Coverage | 13 unique codes used to categorize bill type (not bill theme) via Peltzman (1984). |
| **CRS Policy Area** | 28% Coverage | 32 broad policy areas that can be assigned to bills or resolutions. CRS stands for Congressional Research Service. |
| **Issue Codes** | 55% Coverage | 98 unique codes that provided the most robust option for bill topic tagging, however some topics appeared less relevant for contemporary bills (e.g., Cold War). |

# Appendix B

## Codebook / Data Dictionary

1. **(1) Physical Features and Static Environment**
    - 1.1. Animals and Plants
        - 1.1.1. Crops https://nassgeodata.gmu.edu/CropScape/ (2024; Geolevel: Raster)
    - 1.2. Water
        - 1.2.1. Coastlines https://catalog.data.gov/dataset/tiger-line-shapefile-2019-nation-u-s-coastline-national-shapefile (Geolevel: Vector data polygons)
        - 1.2.2. Wetlands
        - 1.2.3. Bathymetry https://apps.nationalmap.gov/downloader/ (Geolevel: Contour lines)
    - 1.3. Mountains https://apps.nationalmap.gov/downloader/ (Geolevel: Raster, Contour lines)

1.4. Protected Areas https://www.protectedplanet.net/en/thematic-areas/wdpa?tab=WDPA (Year 2024; Geolevel: NA; Worldwide data, need clip)

1.5. Climate

1.5.1. Snowfall
https://www.arcgis.com/apps/mapviewer/index.html?url=https://gis.ncdc.noaa.gov/arcgis/rest/services/snowfall/rsi/MapServer&source=sd (Geolevel: Point; CSV with Longitude and Latitude)
https://noaa-ghcn-pds.s3.amazonaws.com/index.html

1.6. Sinkhole Susceptibility https://www.sciencebase.gov/catalog/item/646d9383d34ee02593fb546f (Raster and Vector)

**2. (2) Infrastructure**

2.1. Bridges https://geodata.bts.gov/datasets/usdot::national-bridge-inventory/about (2024)

2.2. Major Highways https://geodata.bts.gov/datasets/usdot::north-american-roads/about Year: 2025; Line data

2.3. Dams https://nid.sec.usace.army.mil/#/downloads (2025)

2.4. Ports https://geodata.bts.gov/datasets/usdot::principal-ports/about (2025)

2.5. Housing Stock

2.6. Institutions

2.6.1. Military Bases https://data-usdot.opendata.arcgis.com/datasets/usdot::military-bases/about/ (2024)(boundaries)

2.7. Border Crossing Entry Points - https://opendata.sandag.org/Land-and-People-/Border-Crossing-Entry-Data/fcbb-53ng/about_data

2.8. Energy https://atlas.eia.gov/apps/all-energy-infrastructure-and-resources/explore (Geolevel: Point, Year: 2025)

2.9. Transportation and Public Works

2.9.1. Expenditures

2.9.2. Ridership and Modality

**3. (3) Demographics (Static values)**

3.1. Basic Data

3.1.1. Population Size
https://data.census.gov/table/ACSDP1Y2023.DP05?q=population&g=010XX00US$5000000&y=2023
https://www.fns.usda.gov/data-research/data-visualization/snap-community-characteristics-congressional-district-dashboard (2023: Geolevel: CD), Pop density (Imputed)

3.1.2. Race
https://data.census.gov/table/ACSDP1Y2023.DP05?q=population&g=010XX00US$5000000&y=2023 (2023; Geolevel: CD)

3.1.3. Gender and age
https://data.census.gov/table/ACSST1Y2023.S0101?q=gender&g=010XX00US$5000000 (2023; Geolevel: CD)

3.2. Veterans https://data.census.gov/table?q=Veterans&g=010XX00US$5000000&y=2023 S2101 (2023; Geolevel: CD)

3.3. Disability Status https://data.census.gov/table?q=disability&g=010XX00US$5000000&y=2023 S1810 (2023; Geolevel: CD)
https://www.fns.usda.gov/data-research/data-visualization/snap-community-characteristics-congressional-district-dashboard (2023: Geolevel: CD)
By age and ethnicity

3.4. Income SOI Tax Stats - Data by congressional district | Internal Revenue Service (2022; Geolevel: CD)

3.4.1. Income and population, median income, mean income: https://data.census.gov/table?q=income&g=010XX00US$5000000&y=2023 S1901 (2023; Geolevel: CD)

3.4.2. Median HH Income by SNAP: https://www.fns.usda.gov/data-research/data-visualization/snap-community-characteristics-congressional-district-dashboard (2023: Geolevel: CD)

3.4.3. Percent Poverty (%s also broken down by race) https://www.fns.usda.gov/data-research/data-visualization/snap-community-characteristics-congressional-district-dashboard (2023: Geolevel: CD)
https://data.census.gov/table?q=Poverty&g=010XX00US$5000000&y=2023 % can be calculated with population by CD, S1701 (2023: Geolevel: CD)

3.5. Native Americans

3.5.1. Reservations
https://epa.maps.arcgis.com/home/item.html?id=b04b580b3e5e40cc8152b14583814073 (maybe) (2025;Geolevel;National)
https://catalog.data.gov/dataset/tiger-line-shapefile-2020-nation-u-s-american-indian-tribal-subdivisions (2020; Census; Geolevel: NA)

3.5.2. Number of Native Americans
https://data.census.gov/table?q=American+Indian+and+Alaska+Native&g=040XX00US01$5000000&y=2023 DP05 (2023; Geolevel: CD)

3.6. Mobility (Moved recently)
https://data.census.gov/table?q=Residential+Mobility&g=010XX00US$5000000 S0701 (2023; Geolevel: CD)

3.7. Foreign Born https://data.census.gov/table?q=Residential+Mobility&g=010XX00US$5000000 S0201 (2023; Geolevel: CD)

**4. (4) Jobs, Economy and Institutions**

4.1. Agriculture
nass.usda.gov/Publications/AgCensus/2022/Online_Resources/Congressional_District_Profiles/ (2022; Geolevel:CD)

4.2. Business, Commerce, and Economic Development

4.2.1. https://github.com/EIG-Research/EIG-Great-Transfer-Mation

4.3. Labor and Employment https://www.census.gov/naics/ (Geolevel: Tract)

4.3.1. Employment - In labor force
https://data.census.gov/table?q=Employment&g=010XX00US$0400000&d=ACS+1-Year+Estimates+Selected+Population+Profiles (2023, Geolevel: CD)

4.4. Tourism and Recreation and Natural Preserved Lands

4.4.1. National Park: https://public-nps.opendata.arcgis.com/maps/c8d60ffcbf5c4030a17762fe10e81c6a/about Year: (2024; Geolevel: NA)

4.4.2. Parks: https://www.tpl.org/park-data-downloads Year: 2024; Geolevel: NA

4.5. Arts, Culture, Religion, Science

**5. (5) Health and Wellbeing**

5.1. Disease Prevalence

5.1.1. Diabetes: https://data.cdc.gov/500-Cities-Places/PLACES-Local-Data-for-Better-Health-County-Data-20/swc5-untb/about_data Year: 2024; Geolevel: Census Tract

5.1.2. Arthritis: https://data.cdc.gov/500-Cities-Places/PLACES-Census-Tract-Data-GIS-Friendly-Format-2024-/yjkw-uj5s/about_data (Year: 2024; Geolevel: Census Tract)

- 5.1.3. HIV/AIDS: https://aidsvu.org/resources/#/datasets (Year: 2023, Geolevel: County) https://gis.cdc.gov/grasp/nchhstpatlas/tables.html
- 5.1.4. High Blood Pressure: https://data.cdc.gov/500-Cities-Places/PLACES-Census-Tract-Data-GIS-Friendly-Format-2024-/yjkw-uj5s/about_data (Year 2024; Geolevel: Census Tract)

5.2. Obesity: https://data.cdc.gov/500-Cities-Places/PLACES-Local-Data-for-Better-Health-County-Data-20/swc5-untb/about_data (Year: 2024; Geolevel: Tract)

5.3. Incarceration: https://github.com/vera-institute/incarceration-trends (Year: 2023; Geolevel: County)

5.4. Hospital beds: https://www.openicpsr.org/openicpsr/project/222901/version/V1/view?path=/openicpsr/222901/fcr:versions/V1.1/nanda_hospitals_2023_csv.zip&type=file (Year: 2023; Geolevel: Tract)

5.5. Medicare/Medicaid:
- 5.5.1. Medicarehttps://data.cms.gov/summary-statistics-on-use-and-payments/medicare-geographic-comparisons/medicare-geographic-variation-by-national-state-county (Year: 2023; Geolevel: County)
- 5.5.2. Lack of Health Insurance: https://data.cdc.gov/500-Cities-Places/PLACES-Census-Tract-Data-GIS-Friendly-Format-2024-/yjkw-uj5s/about_data (Year 2024; Geolevel: Tract)

5.6. Clinics
- 5.6.1. Pregnancy Centers

5.7. SNAP/EBT/WIC/FARMS (free and reduced meals) Food Assistance - https://data.census.gov/table/ACSST1Y2023.S2201?q=snap&g=010XX00US$5000000 (ASC 2023 - geolevel: CD) https://www.fns.usda.gov/data-research/data-visualization/snap-community-characteristics-congressional-district-dashboard (2023: Geolevel: CD)

5.8. Drinking: https://data.cdc.gov/500-Cities-Places/PLACES-Census-Tract-Data-GIS-Friendly-Format-2024-/yjkw-uj5s/about_data (Year 2024; Geolevel: Tract)

**6. (6) Dynamic Events and Behaviors**

6.1. Hazards (FEMA) https://www.fema.gov/about/openfema/data-sets and https://github.com/FEMA/openfema-samples (Floods, fire declarations)

6.2. Traffic
- 6.2.1. Commuters https://github.com/mjharris95/Science-Impacts/blob/main/data/OD_countySum001_2016.csv (Year: 2016; Geolevel: CD (original data from Census LODES-LEHD)). Data includes commuters for different sectors (e.g., mining sector, education sector, federal jobs).

6.3. Pollution
- 6.3.1. AQI https://www.epa.gov/outdoor-air-quality-data (Year: 2NAICU - Federal Student Aid Data Sheets 024; Geolevel: County)

6.4. War events / attacks, Festivals \ Sporting Events \ The Olympics, Infestations

**7. (7) Education**

7.1. University funding Awards and total dollar amounts, broken out by Pell Grants, campus-based programs (SEOG, Federal Work-Study, and Perkins Loans), and federal loan programs. https://www.naicu.edu/policy-advocacy/federal-student-aid-data-sheets/ (Geolevel: CDs, but data is in distributed PDFS)

7.2. Education Funding https://nces.ed.gov/ccd/elsi/tablegenerator.aspx(Geolevel: CD)

7.3. K-12 -https://nces.ed.gov/ccd/elsi/tablegenerator.aspx

7.4. Education attainment: https://data.census.gov/table?q=Education&g=010XX00US$0400000&d=ACS+1-Year+Estimates+Selected+Population+Profiles S0201 (2023, Geolevel: CD)

7.5. Special education

**8. (8) Other / MISC**

8.1. Vulnerable Populations and Social Vulnerability https://www.atsdr.cdc.gov/place-health/php/svi/svi-data-documentation-download.html (Geolevel: Tract)
8.2. Social Capital https://dataforgood.facebook.com/dfg/tools/social-capital-atlas ( Geolevel: County)
8.3. Social Capital [add state college data] (Geolevel: County)
8.4. Losses due to NIH Overhead Cuts and Federal Job losses in Science https://github.com/mjharris95/Science-Impacts/blob/main/output/NIH_impact_cong.csv (Year 2025; Geolevel: CD)
8.5. Urban vs. Rural https://www.openicpsr.org/openicpsr/project/110663/version/V1/view (Year 2016; Geolevel: Tract)